\documentclass[fleqn,usenatbib]{mnras}
\usepackage{newtxtext,newtxmath}
\usepackage{hyperref} 
\usepackage{mathtools}
\usepackage{graphics}
\usepackage{float} 
\usepackage{booktabs}
\usepackage{xcolor}
\usepackage[T1]{fontenc}
\usepackage{graphicx}	
\usepackage{amsmath}

\title[Bending instabilities of $m = 1$ mode]{Bending instabilities of $m = 1$ mode in disc galaxies: interplay between dark matter halo and vertical pressure}

\author[S Goyary et al.]
         {Sagar S. Goyary$^{1}$\thanks{E-mail: sagar.singhgoyary@rgu.ac.in / heisnam.singh@rgu.ac.in} 
        Kanak Saha$^{2}$\thanks{E-mail: : kanak@iucaa.in},
        H. Shanjit Singh$^{1}$ and Suchira Sarkar$^{2}$
\\
$^{1}$  Department Of Physics, Rajiv Gandhi University, Arunachal Pradesh-791112, India\\
$^{2}$  Inter-University Center for Astronomy and Astrophysics, Pune-411007, India}

\date{Accepted XXX. Received YYY; in original form YYY}

\pubyear{2023}

\begin{document}
\label{firstpage}
\pagerange{\pageref{firstpage}--\pageref{lastpage}}
\maketitle

\begin{abstract}
A self-gravitating, differentially rotating galactic disc under vertical hydrostatic equilibrium is supported by the vertical pressure gradient force against the gravitational collapse. Such discs are known to support various bending modes e.g., warps, corrugation, or scalloping (typically, higher order bending modes) of which $m=1$ bending modes (warps) are the most prevalent ones in galactic discs. Here, we present a detailed theoretical analysis of the bending instability in realistic models of disc galaxies in which an exponential stellar disc is under vertical equilibrium and residing in a cold rigid dark matter halo. A quadratic eigenvalue equation describing the bending modes is formulated and solved for the complete eigen spectrum for a set of model disc galaxies by varying their physical properties such as disc scale-height, and dark matter halo mass. It is shown that the vertical pressure gradient force can excite unstable bending modes in such a disc as well as large scale discrete modes. Further, it is shown that the unstable eigen-modes in a thinner disc grow faster than those in a thicker disc. The bending instabilities are found to be suppressed in discs dominated by massive dark matter halo. We estimate the growth timescales and corresponding wavelength of the $m=1$ unstable bending modes in Milky Way like galaxies and discuss its implication.

\end{abstract}

\begin{keywords}instabilities - galaxies: kinematics and dynamics
 - galaxies: structure - methods: analytical
\end{keywords}

\section{Introduction}
\label{sec:intro}

Large-scale bending waves or the classic integral-sign warps ($m=1$, where $m$ is the azimuthal wave number) are commonly seen in nearby disc galaxies including our Milky Way \citep{1957AJ.....62...90B, 1957AJ.....62...93K, van1979optical, 1976A&A....53..159S, bosma1991warped, Levine+2006, 2019NatAs...3..320C, Cheng+2020, 2021ApJ...912..130C}. More than 50\% of local disc galaxies show warping of their disc mid-plane when viewed edge-on \citep{1990Ap&SS.171..239S, 1998A&A...337....9R, 2003A&A...399..457S, AnnPark2006}. 
Considering the fact that the detection of warps is a tedious task in low-inclination discs - most disc galaxies may be warped - simply speaking galactic discs refuse to stay flat \citep{Lynden-Bell1965, HunterToomre1969}. Although gaseous discs exhibit large amplitude bending waves, typically $\sim 1$~kpc size \citep{1982ApJ...259L..63K, Levine+2006}, they are of small amplitude in the stellar discs; typically not more than a few hundred parsecs \citep{Sahaetal2009,2020PASJ...72...53S, 2020MNRAS.495.3705N} - adding woes to the detection of stellar bending waves further in external galaxies. However, there is a sudden rejuvenation in this age-old topic with the availability of RAVE (Radial Velocity Experiment) survey covering more than a few kpc distances from the solar neighborhood in the Milky Way \citep{2006AJ....132.1645S}; 6D phase space data combining recent surveys like GAIA, LAMOST, SEGUE \citep{2009AJ....137.4377Y, 2013MNRAS.436..101W, 2014MNRAS.439.1231B, 2018A&A...616A..11G, wang2020mapping} to decode the dynamical fossil signature of bending modes in our Galaxy \citep{widrow2012galactoseismology,widrow2014bending, xu2015rings, 2018MNRAS.478.3809S, 2018Natur.561..360A, 2018MNRAS.481L..21P, bennett2019vertical, 2022MNRAS.513.3065N}. In other words, bending waves are present both in stars and gas - implying that they might be long-lived and rather easier to be generated in their host discs \citep{Gomez+2013, ChequersWidrow2017, Chequers+2018}.

Various mechanisms are proposed to generate bending waves in galaxies. For example, tidal interaction with a companion galaxy or satellite has been suggested to generate large-scale bending waves in the Galaxy and in external galaxies \citep{HunterToomre1969, 1997ApJ...480..503H, 2006ApJ...641L..33W,saha2006generation, DubinskiChakrabarty2009}. A number of studies focused on the infalling gas from the intergalactic medium, cosmic infall  \citep{KahnWoltjer1959, JiangBinney1999,  LopezCorredoiraetal2002, ShenSellwood2006, Sanchez-Salcedo2006} as well as misaligned disc and dark matter halo \citep{DebattistaSellwood1999} to generate large scale bending of the galactic discs. Although large-scale bending waves can be generated through these processes, their persistence remains a perennial problem. Substantial efforts have been invested over the last few decades to generate long-lived bending waves via internal instabilities \citep{HunterToomre1969, Araki1985, 1984sv...bookQ....F, Sparke1995, Sellwood1996, RevazPfenniger2004, ChequersWidrow2017} including counter-rotation \citep{1994ApJ...425..530S, 1996MNRAS.283..543K, 2022MNRAS.516.3175B} and dynamical friction due to the dark matter halo or its substructure \citep{NelsonTremaine1995,WeinbergBlitz2006} as the external effects might not always be present. The readers are referred to \cite{binney1992warps} for a detailed account on disc warping.

\cite{HunterToomre1969} in their pioneering work, found no unstable bending modes in a family of rotating, self-gravitating razor-thin cold discs and concluded that discrete bending modes can not account for the observed gentle warping of a disc. However, this analysis needs to be carried out in a more realistic model of warm disc galaxies that includes additionally a dark matter halo. A detailed study of the linear normal mode analysis carried out by \cite{Araki1985} on a thin stellar slab with uniform density distribution along the radial axis concludes that when the ratio of vertical to radial velocity dispersion of stars $\sigma_z/\sigma_R \leq 0.293$, the slab becomes bending unstable. This has been confirmed later on by various authors \citep{Sellwood1996, 2005AstL...31...15S, 2010AN....331..731K, 2017A&A...597A.103K, 2011MNRAS.415.1259G, 2013MNRAS.434.2373R} through analytical studies as well as N-body simulations.  \cite{Sellwood1996}, in particular, carried out a detailed analysis of bending instability in a more realistic model of discs - a family of self-gravitating Kuzmin-Toomre discs with finite thickness and found that they exhibit both discrete and exponentially growing modes which disappear as the disk thickness increases. He further demonstrated using N-body simulations that one of the disk models (KT/5) supports long-lived discrete axisymmetric ($m=0$) bending modes. Discrete normal modes of oscillation of a galactic disc are also shown to exist in a number of studies but for specific galactic systems \citep{Mathur1990, Weinberg1991, Louis1992, MillerSmith1994}. In other words, long-lived bending oscillations pose a considerable challenge to dynamicists; the readers are referred to \cite{binney1992warps} for an insightful discussion in this context. It is in this spirit, we revisit this fascinating problem of galactic dynamics and extend the analysis to realistic models of disc galaxies considering an exponential density distribution along the radial direction and sech$^2$ or exponential along the vertical direction \citep{1942ApJ....95..329S, freeman1970disks, 1988A&A...192..117V, 2023arXiv230304171H} and a logarithmic dark matter halo potential that gives rise to a flat rotation curve \citep{2008gady.book.....B}. The rest of the paper presents a detailed analytic study of the $m=1$ bending instability, the nature of the eigen spectrum, and how the growth rate of bending instability depends on various properties of the galaxy including that of the dark matter halo.

This paper is organized as follows: In Section \ref{sec2}, we give a description of the dynamical model of the disc bending and the details of the galaxy and dark matter halo density profiles. In Section \ref{sec3}, the methods used for solving the numerical solution of models are presented with input parameters to solve the quadratic eigenvalue problem and find the nature of eigenmodes. The results of the numerical analysis of eigen modes  in the absence and presence of vertical pressure  are discussed in Section \ref{sec4} and \ref{sec5} respectively. In Section \ref{sec:WKB}, we estimate and  discuss the growth time of bending mode  and wavelength using the WKB dispersion relation. In Section \ref{sec7}, we discuss our results and present the conclusions.

\section{Formulation of the problem} \label{sec2}
\subsection{Model of the galactic disc and dark matter halo}

We use galactocentric cylindrical coordinate system ($R$, $\mathrm{\phi}$, $z$). We consider the density distribution of the galactic disc  exponential out to truncation radius $R_t$ and is zero beyond a radius $R_o$ ($<R$), and between these two radii, the density tapers smoothly with $\cos^2$ function along the radial direction  introduced by \cite{1988MNRAS.234..873S}  and Gaussian in the vertical direction, as given by 

\begin{flalign}
     \rho (R,z) &=  \rho_{_{0,0}} \text{e}^{-\frac{R}{R_d}}\text{e}^{-\frac{z^2}{z_{_0}^2}},~~~\text{if } ~~R\leq R_t \notag\\
    &=  \rho_{_{0,0}}\text{e}^{-\frac{R}{R_d}} \text{e}^{-\frac{z^2}{z_{_0}^2}}  \cos^2\left( \frac{\pi}{2} \frac{R-R_t}{R_o-R_t} \right),~~~~ \text{if }~~R_t\leq R\leq R_o \notag\\
    &=  0, ~~~~ \text{if } ~~R\geq R_o \label{eqn:disc_density}
\end{flalign}

\noindent Here $\rho_{_{0,0}}$ is the central, mid-plane ($z=0$) density value of the galactic disc, given by $\rho_{_{0,0}}= \frac{M_d}{2\pi^{3/2} R_d^2 z_{_0}}$.  $M_d$ is the mass of the disc, $R_{d}$ denotes radial scale length and $z_{0}$ denotes the vertical scale height. The potential corresponding to the above density distribution is calculated in the following.

\begin{equation}
\begin{split}
    \Phi_{disc}(R,z) = - 2 \pi^{3/2} G \rho_0 R_d^2 z_0\int^\infty_0 dk J_0(kR) I(R,k) \times I(z,k),
\end{split}
\end{equation}
 where  $I(R,k)$ and $I(z,k)$  are given by
\begin{equation*}
\begin{split}
     I(R,k) & = \int^{R_t}_0 R'dR' J_0(kR') e^{-R'/R_d} \\
      &+ \int^{R_o}_{R_t} R'dR' J_0(kR') e^{-R'/R_d} \cos^2{\left(\frac{\pi}{2}\frac{R'-R_t}{R_o-R_t}\right)},
\end{split}
\end{equation*}

 \begin{equation*}
 \begin{split}
     I(z,k) &= \frac{1}{2} \exp \left( \frac{k^2z_0^2}{4}-kz\right) \text{erfc}\left( \frac{kz_0}{2}-\frac{z}{z_0}\right)\\
    & +\frac{1}{2} \exp \left( \frac{k^2z_0^2}{4}+kz\right) \text{erfc}\left( \frac{kz_0}{2}+\frac{z}{z_0}\right).
    \end{split}
 \end{equation*}
At the mid-plane of the galactic disc $z=0$,
 \begin{equation*}
     I(k)=  \exp \left( \frac{k^2z_0^2}{4}\right) \text{erfc}\left( \frac{kz_0}{2}\right).
 \end{equation*}
\noindent Here $J_0(kR)$ is the cylindrical Bessel function of the first kind of order zero. The circular velocity due to the disc, ${V_{c,d}}$, at any radius $R$, and at the disc mid-plane is obtained  by using the following relation \citep{2008gady.book.....B}

\begin{equation}
    V^2_{c,d} =  R \frac{\partial\Phi_{disc}}{\partial R}\bigg|_{z=0}.
\end{equation}
We consider the density distribution  of the dark matter halo of the form \citep{2008gady.book.....B}

\begin{equation*}
    \displaystyle\rho_{h}(R,z)=\frac{V_{0}^2}{4\pi Gq^2}\frac{(2q^2+1)R_c^2+R^2+(2-q^{-2})z^2}{(R_c^2+R^2+z^2q^{-2})^2},
\end{equation*}
where $R_c$ is the core radius and $q$ is the halo flattening parameter, i.e. the axis ratio of the equipotential surfaces. 
$V_{0}= \sqrt{\frac{GM_h}{R_c}}$ is the flat circular speed at large $R$. $M_h$ denotes dark matter halo mass. The corresponding gravitational potential of the dark matter halo is given by:

\begin{equation*}
\Phi_{halo}(R,z)= \frac{V^2_{0}}{2}\ln\bigg(R^2+R_c^2+\frac{z^2}{q^2}\bigg).
\end{equation*}
    
\noindent The circular velocity $V_{c,h}(R)$ at radius $R$ on the equatorial plane of the halo potential is given by \citep{2008gady.book.....B}

\begin{equation*}
    V_{c,h}(R) = \frac{V_{0}R}{\sqrt{R^{2}+R_{c}^{2}}}.
\end{equation*}

\begin{figure}
    \centering
\includegraphics[width=.85\columnwidth]{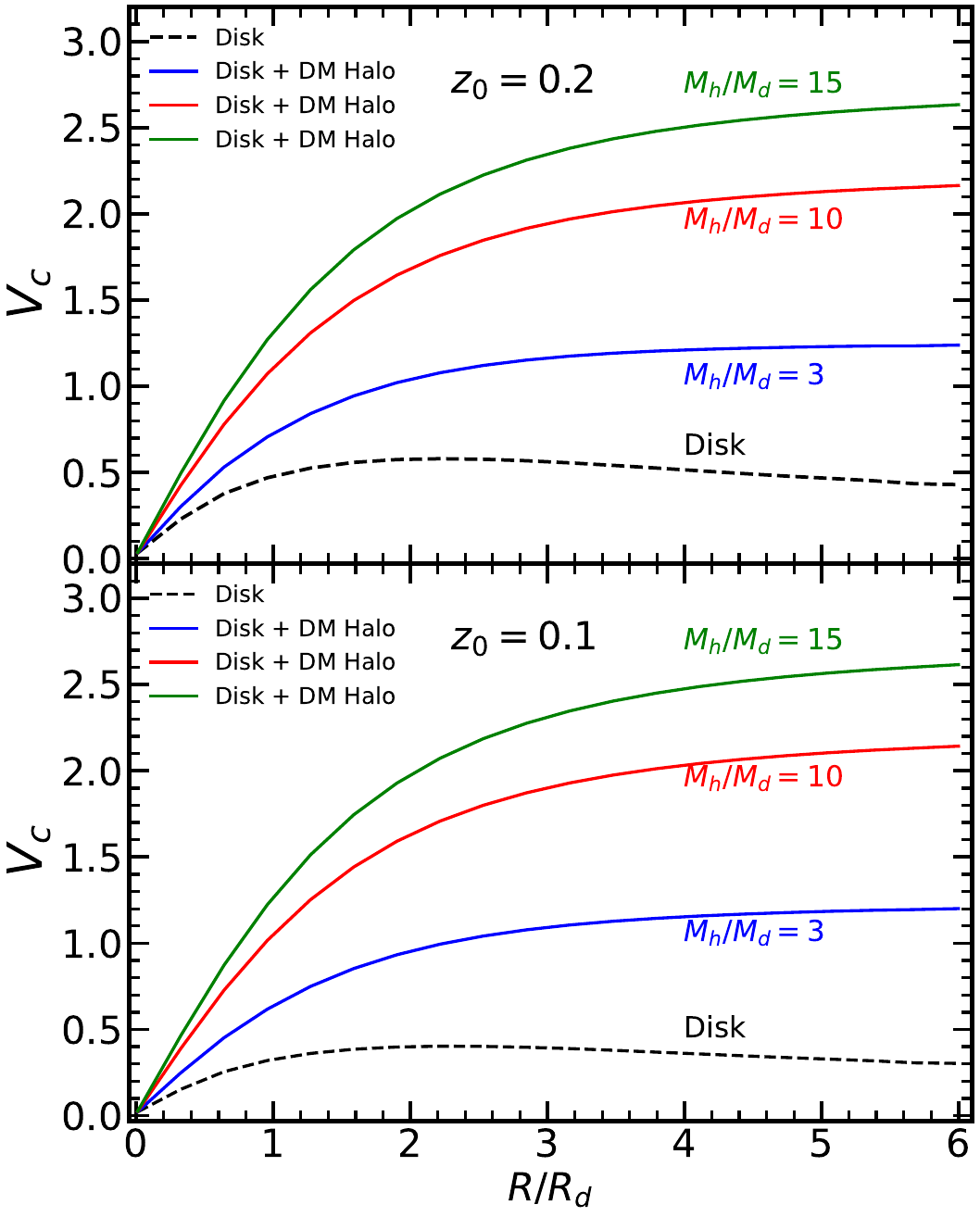}
\caption{\textbf{Rotation curve:}  Circular velocity ($V_{c}$) curves for the galactic disc and the disc+ dark matter halo. The top panel shows the velocity curves when the vertical scale height $z_{0}$ of the disc is $0.2R_{d}$, and the bottom panel shows the curves when $z_{0}=0.1R_{d}$. In each panel we show the velocity curves for the disc+halo case for three values of halo to disc mass ratio, taking $M_{h}/M_{d}$ to be 3, 10, and 15. The velocities are normalised to $\sqrt{GM_d/R_d}$, where $G =M_{d} = R_{d} =1$ is considered (see Section \ref{sec3}).}
\label{fig:R_C}
\end{figure}

\noindent This profile yields an asymptotically flat rotation curve as observed in local spiral galaxies through the 21cm HI observation \citep{Bosma1981, Walteretal2008, Swatersetal2009}. The net circular velocity is obtained by adding the contribution of the disc and the halo in quadrature as $V_{c, total}=(V^{2}_{c,d}+V^{2}_{c,h})^{1/2}$. We show the circular velocities for the disc and the disc plus halo cases, corresponding to different values of $z_{0}$ and the halo to disc mass ratio ($M_{h}/M_{d}$), in Fig.\ref{fig:R_C}.

\subsection{Derivation of the dynamical equation}
\label{sec:bendingequation}

We consider an axisymmetric galactic disc of density profile given by equation (\ref{eqn:disc_density}), rotating in the equatorial plane ($z=0$) of the dark matter halo with an angular speed $\Omega(R)$ about the symmetry axis of the halo ($R=0$). The halo is considered to be rigid or non-responsive throughout the analysis. Then the dynamical equation of the bending of the galactic disc under small angle approximation (i.e., considering vertical displacement w.r.t. to unperturbed plane $z=0$ to be small) can be written as: 

\begin{equation}
    \left(\frac{\partial}{\partial t} + \Omega(R)\frac{\partial}{\partial \phi} \right)^2 Z = -{\nabla_{z}}{\Phi_{self}} - \nabla_{z}{\Phi_{halo}} -\frac{1}{\rho_{d}} \nabla_{z}{P},
    \label{eqn:dynamic}
\end{equation}

\noindent where $Z$ denotes the small bending above the disc mid-plane; $\nabla_{z}$ is the vertical gradient;  the first term on the RHS is the force due to the self-gravity, the second term is the vertical restoring force due to the dark matter halo and the third term is the vertical pressure force due to the non-zero vertical velocity dispersion respectively. The system ignores any diffusion of matter along the radial direction, such as might be caused by epicyclic motions in the stellar disc. The dynamical model of the bending of the disc adopted here is similar to that of \cite{1988MNRAS.234..873S} and \cite{saha2006generation}, with the additional vertical pressure gradient term introduced here.  

The vertical force $F_{d}=-{\nabla_{z}}{\Phi_{self}}$ due to the disc self-gravity is given by

\begin{flalign}
		F_{d} = -G\int_{0}^{\infty} &\int_{-\infty}^{\infty}\int_{0}^{2\pi}R'\rho(R',z')\notag \\
          &\times\frac{\left[Z_m(R,\phi,t)-Z_m(R',\phi',t')\right]d\phi'dz'dR'}{\left[R^2+R'^2-2RR'\cos(\phi-\phi')+(z-z')^2\right]^{\frac{3}{2}}}.\label{eqn:disc_force}
\end{flalign}

\noindent Here $m$ represents $m ^{th}$ order bending mode. 

The vertical restoring force near the disc plane due to the dark matter halo is given by  

\begin{flalign}
		F^m_{\text{h}} &= -\nu_{\text{h}}^2(R)Z_m(R,\phi,t), \label{eqn:halo_force}
\end{flalign}
where $\nu_{\text{h}}$ is vertical frequency due to the dark matter halo alone in the unperturbed disc plane. When the vertical frequency is greater than the orbital frequency  i.e. $\nu^2(R)>\Omega^2(R)$, the disc oscillates in the vertical direction more rapidly than it orbits about the galactic center \citep{1984ApJ...280..117S}. 

\begin{figure*}
\centering
\includegraphics[width=.85\columnwidth]{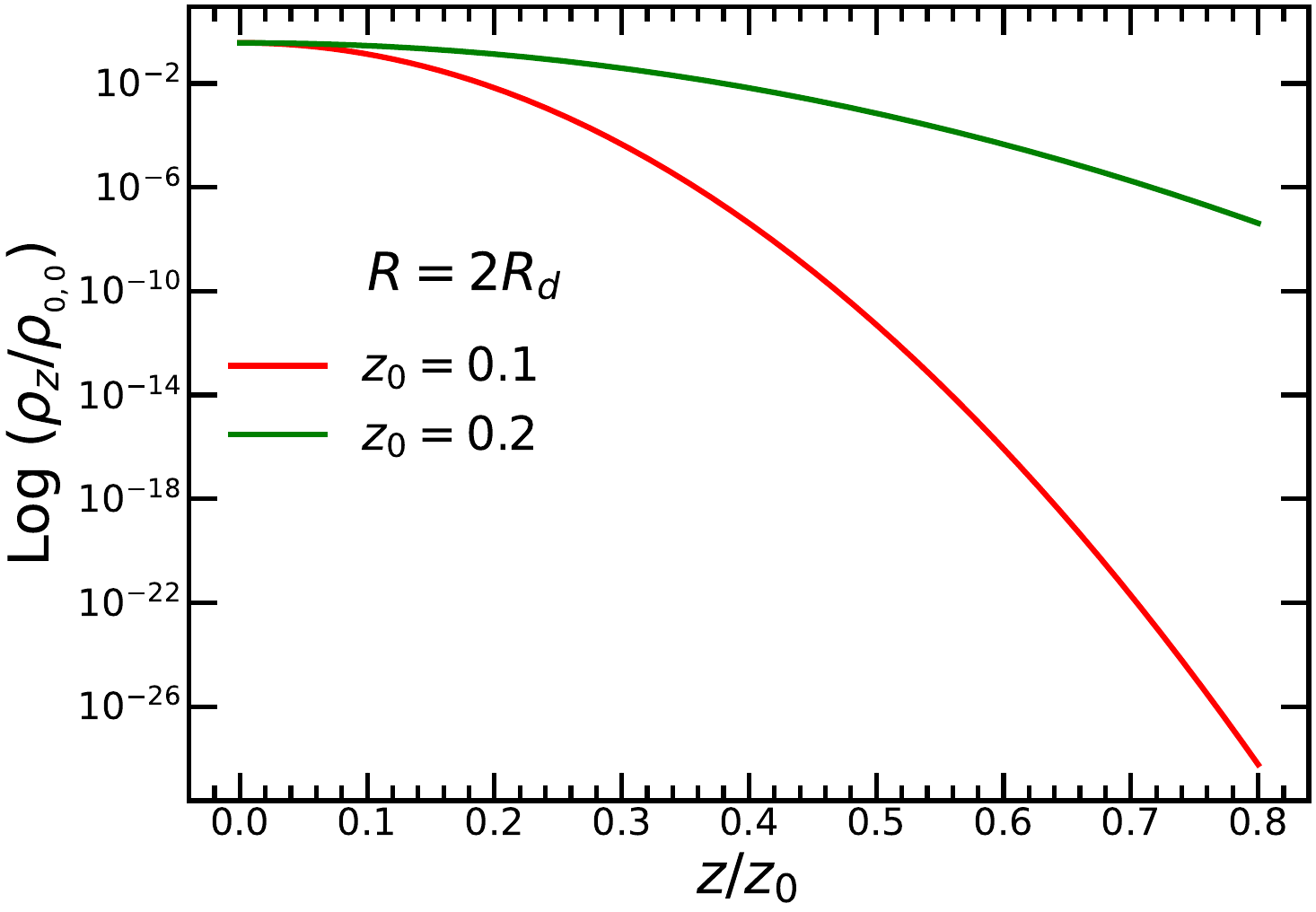}  \hspace{1cm}\includegraphics[width=.8\columnwidth]{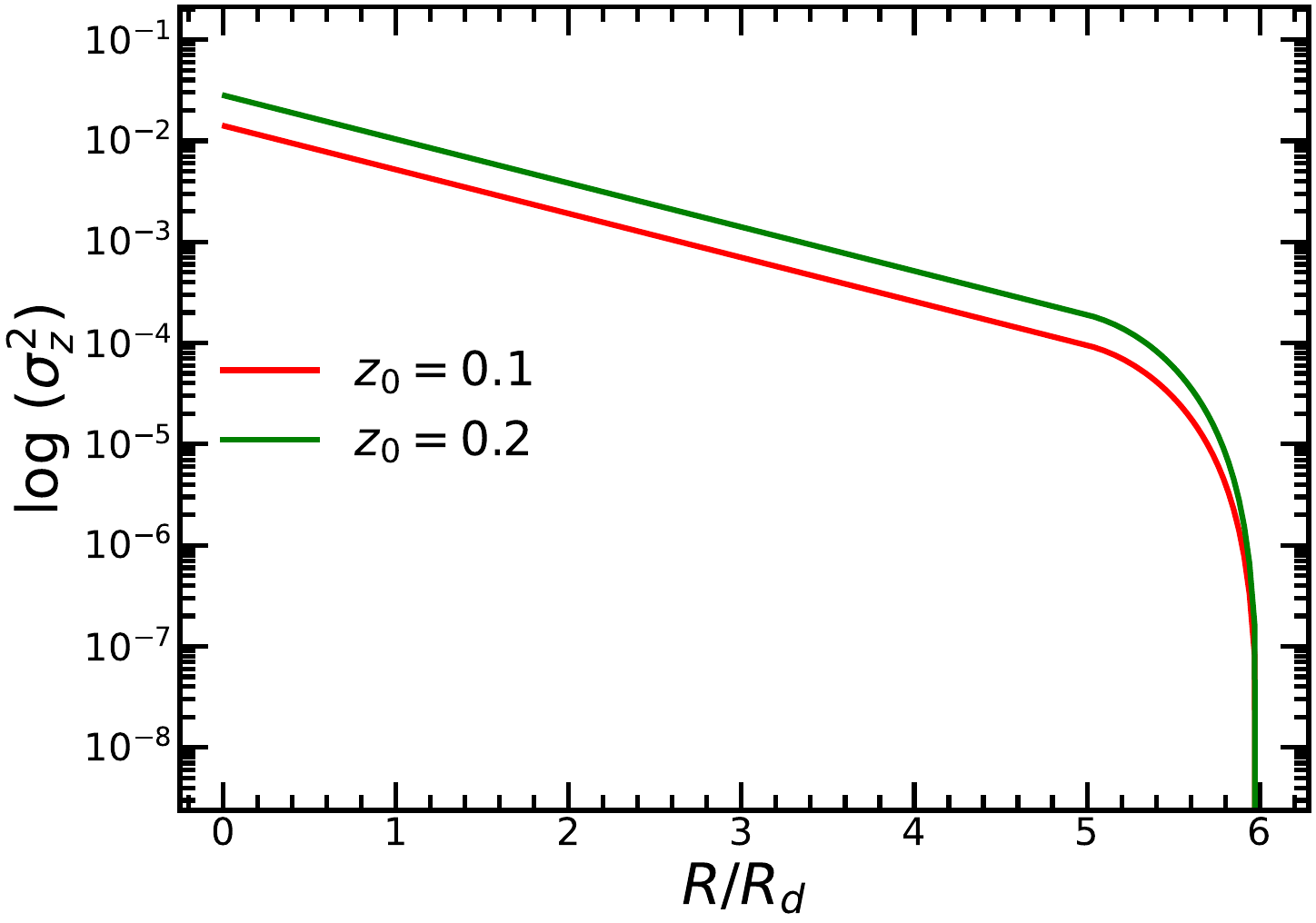}
    \caption{\textbf{Density and velocity dispersion:} Vertical  density distribution $\rho_z$  and  vertical velocity dispersion $\sigma_z$ of the disc. The left panel shows the vertical  density distribution $\log (\rho_z/\rho_{_{0,0}})$ as a function of the vertical disc height $z/z_0$ for  $z_0=0.1 $ and 0.2 at radial distance $R = 2 R_d$. The right panel shows the vertical velocity dispersion, $\sigma_z$ in log scale  as a function of the radial distance for two different vertical disc scales $z_0 =0.1$ and  $0.2$  respectively. The velocity dispersion $\sigma_z$ is in the unit of $\sqrt{GM_d/R_d}$.}
    \label{fig:Density-Dispersion}
\end{figure*}

\subsubsection{Calculation of the vertical velocity dispersion}
\label{sec:verticalpressure}

Assuming barotropic stellar fluid, the pressure $P$ acting along the vertical direction arising from the non-zero vertical velocity dispersion in the disc is given by

\begin{align*}
	P &=  \sigma^2_z(R,z) \rho(R,z), \\
	   &= \sigma^{2}_{z}(R,z)\rho_{_{0,0}} \rho (R) e^{-\frac{z^2}{z_0^2}  }.
\end{align*}

\noindent In the above equation $\sigma_z(R,z)$ denotes the vertical velocity dispersion. Now the Poisson equation for an axisymmetric, thin galactic disc is given by 

\begin{equation}
    \frac{dK_z}{dz} = -  4 \pi G \rho(R,z),
\end{equation}
 
\noindent where $K_{z}$ denotes the vertical force per unit mass. Substituting $\rho(R,z)$ into the above equation and integrating the equation from $z=0$ to $z$, we obtain

 \begin{equation*}
    \begin{split}
        K_z = -2 \pi^{3/2} G \rho_{_{0,0}}   z_0  \rho(R) \text{erf}\left(\frac{z}{z_0}\right).
    \end{split}
 \end{equation*}
 
\noindent Now, the condition for the vertical hydrostatic equilibrium \citep{1977LNP....69.....R} for the disc is given by

\begin{equation}
     \frac{\partial}{\partial z}\bigg(\rho(R,z) \sigma^2_z(R,z)\bigg)  = \rho(R,z) K_z.
\end{equation}

\noindent We obtained an analytical expression for $\sigma_{z}$ by integrating the above equation from $z$ to $\infty$, with the boundary condition of $\rho(z)=0$ as z $\to$ $\infty$ \citep{2017ApJ...836..181D}. This gives us 

\begin{equation*}
\begin{split}
    \sigma^2_z(R,z) & = -\frac{1}{\rho(R,z)} \int_{z}^{\infty} \rho(R,z')dz' K_{z'}, \\
    & =  2 \pi^{3/2} G  \rho_{_{0,0}} \rho(R)  z_0 e^{\frac{z^2}{z_0^2}} \int_{z}^{\infty} e^{-\frac{z'^2}{z_0^2}
    }\text{erf}\left(\frac{z'}{z_0}\right) dz'. 
\end{split}
\end{equation*}
At the mid-plane of the disc $z=0$,

\begin{equation}
\begin{split}
    \sigma^2_z(R,0) & =  2 \pi^{3/2} G  \rho_{_{0,0}}  z_{0} \rho(R) \int_{0}^{\infty} e^{-\frac{z'^2}{z_0^2}
    }\text{erf}\left(\frac{z'}{z_0}\right) dz'.\label{eqn:sigma_square1}
\end{split}
\end{equation}

\noindent Using the following integral relation \citep{gradshteyn2014table}-
\begin{equation*}
\begin{split}
    \int_{0}^{\infty} e^{-b^2x^2
    }\text{erf}\left(ax\right) dx& = \frac{\sqrt{\pi}}{2b}-\frac{1}{b\sqrt{\pi}}\tan^{-1}\left(\frac{b}{a}\right),
\end{split}
\end{equation*}

\noindent and substituting $a=b=\frac{1}{z_0}$ we get,

\begin{equation}
\begin{split}
    \int_{0}^{\infty} e^{-\frac{z'^2}{z_0^2}
    }\text{erf}\left(\frac{z'}{z_0}\right) dz'& =  \frac{\pi}{4}z_0.
    \label{eqn:integral_relation}
\end{split}
\end{equation}

\noindent Substituting equation (\ref{eqn:integral_relation}) into equation (\ref{eqn:sigma_square1}) we get the final expression for $\sigma^2(R,0)$ as

\begin{equation}
\begin{split}
    \sigma^2_z(R,0) & =  \frac{\pi^{2} }{2} G  \rho_{_{0,0}} z_0^2 \rho(R). \label{eqn:sigma_square}
\end{split}
\end{equation}

\noindent Equation (\ref{eqn:sigma_square}) provides the vertical velocity dispersion of the disc at $z=0$ as a function of disc radius. The right panel of Fig. \ref{fig:Density-Dispersion} shows the plot of vertical velocity dispersion for two different discs of vertical scale heights $z_0 =0.1$ \&  $z_0 = 0.2$ at radial distance $2R_d$. We see that at any radial distance, the disc of $z_0 = 0.2$ has a higher $\sigma_z$ value than the disc of $z_0=0.1$. The velocity dispersion sharply drops in between truncation radius $R_t =5$ and $R_o =6$.  In the left panel of Fig.\ref{fig:Density-Dispersion} we show the vertical density distribution of the discs with scale heights $z_0=0.1$ and 0.2 at radial distance 2$R_d$.

\subsubsection{The quadratic eigenvalue problem (QEP)}

When equations (\ref{eqn:disc_force}) and (\ref{eqn:halo_force}) are substituted  into equation (\ref{eqn:dynamic}) we can  get the solution of dynamical equation (\ref{eqn:dynamic} ) in the form of Fourier terms as-
\begin{eqnarray}
    Z_m(R, \phi, t) =  \mathbb{R}\{h(R) e^{j(\omega_m t-m\phi)}\}.
    \label{eqn:Z}
\end{eqnarray} 
Here $m$ represents the $m^{th}$ bending mode of the disc, which represents an $m$-armed warp precessing at eigen frequency $\omega = \omega_R \pm j \omega_I$  (i.e., the whole pattern revolves with period $2\pi/\omega$). $h(R)$  describes   the shape of  bending  modes in the self-gravitating disc.

When the imaginary part of the eigen frequency, i.e. $\omega_{I}<0$ the eigen mode becomes unstable and grows exponentially with a growth rate of $1/|\omega_{I}|$. On the other hand, when $\omega_{I}>0$ the eigen mode is called damping mode. The eigen mode is stable when $\omega_{I}=0$.

Finally we substitute the expression of $Z_{m}(R,\phi,t)$ from equation (\ref{eqn:Z}) into equation (\ref{eqn:dynamic}) and obtain the following equation
\begin{flalign*}
		\left(\frac{\partial}{\partial t} + \Omega(R)\frac{\partial}{\partial\phi}\right)^2Z_{m}& = -G\int_{-\infty}^{\infty} \int_{0}^{\infty}\int_{0}^{2\pi}R'\rho(R')\rho(z')\\
      & \times \frac{[Z_m(R,\phi,t)-Z_m(R',\phi',t)] d\phi'dR'dz'}{\left[R^2+R'^2-2RR'\cos(\phi-\phi')+(z-z')^2\right]^{\frac{3}{2}}} \\ 
		& -\nu_{h}^2(R)Z_m(R,\phi,t)+\frac{2\sigma_z^2(R)}{z_0^2} Z_m(R,\phi,t).
\end{flalign*}
The triple integral on the right side of the above equation can be written as
\begin{equation}
\begin{split}
\bigg[(\omega_m-m\Omega (R))^2-\nu^2_h\bigg]&h(R) =Gh(R)\int_{0}^{\infty} R' \rho(R')dR' K_0(R,R') \\
&-G\int_{0}^{\infty} R' \rho(R')dR' K_m(R,R')  h(R'). \label{eqn:Int eqn}
\end{split}
\end{equation}
Note that the angular speed $\Omega(R)$ in the above equation has a contribution from  the perturbed disc ($\Omega_d$) as well as the dark matter halo ($\Omega_h$), i.e,
\begin{equation*}
    \Omega^2(R) = \Omega_d^2(R)+\Omega_h^2(R).
\end{equation*}

\noindent In equation (\ref{eqn:Int eqn}), $K_{0}(R,R')$ and $K_{m}(R,R')$ are respectively given by 

\begin{flalign*}
	K_0(R,R') & = \int_{-\infty}^{\infty} \rho(z') B_0(R,R')  dz',\\
       K_m (R,R')& = \int_{-\infty}^{\infty} \rho(z') B_m(R,R')  dz',
\end{flalign*}

\noindent where $B_{0}(R,R')$ and $B_{m}(R,R')$ are given by

\begin{flalign*}
	B_0(R,R') & = \int_{0}^{2\pi} \frac{d\psi}{[R^2+R'^2+(z-z')^2-2RR'\cos(\psi)]^{\frac{3}{2}}},\\
	B_m(R,R') & = \int_{0}^{2\pi} \frac{\cos({m\psi})d\psi}{[R^2+R'^2+(z-z')^2-2RR'cos(\psi)]^{\frac{3}{2}}},
\end{flalign*}

\noindent with $\psi=\phi-\phi'$. 

\noindent Now we put,

\begin{flalign*}
	\nu^2_d = G\int_{0}^{\infty}\rho(R')K_0(R,R')R'dR',
\end{flalign*}
in equation (\ref{eqn:Int eqn}), where $\nu_d$ denotes the vertical frequency of the perturbed disc, and we rearrange the equation in the following compact form   
\begin{flalign}
	\left[ (\omega^2_m-2m\Omega(R_i)\omega_m) \right] h_m(R_i) & = \sum_{j=1}^{N}S_{ij} h_m(R_j),
 \label{eqn:w_square}
\end{flalign}
where 
\begin{equation*}
    \begin{split}
        S_{ij} & = U_{ij} + \delta_{ij}\left( \nu^2_h(R_i) +\nu^2_d(R_i)-\Omega^2(R_i) - \frac{2\sigma^2_z(R_i)}{z_0^2} \right), \\
        U_{ij} & = \Delta R G \rho(R_j) K_m(R_i,R_j) R_j.
    \end{split}
\end{equation*}
Recasting equation (\ref{eqn:w_square}) into a matrix-eigenvalue problem on a uniform grid with $N$ radial points we get a quadratic eigenvalue equation, for the $m^{th}$ order mode, in the form similar to \cite{saha2006generation}:

\begin{equation}
    (\omega_m^2I+\omega_mD+S)h_m=0, \label{eqn:QEP}
\end{equation}
where $I$ is an $N\times N$ identity matrix and $D_{ij}=-2m\Omega(R_i)\delta_{ij} $ . Here $h_m$ is the eigenvector  corresponding to the eigenvalue $\omega_m$. The eigenfrequency $\omega_m$ gives the oscillation frequency of the $m^{th}$ bending mode with the shape specified by $h_m$. The matrices; $I$, $D_{ij}$, and  $S_{ij}$ are commonly known as the mass matrix, damping matrix, and  stiffness matrix, and  are $N\times N$ real square matrices. Equation (\ref{eqn:QEP}) represents a class of nonlinear eigenvalue problem and its solution describes the global behaviour of $m=0,1,2,3...$, etc. modes in the self-gravitating disc. 

\begin{figure*}
\centering
\includegraphics[width=.6\columnwidth]{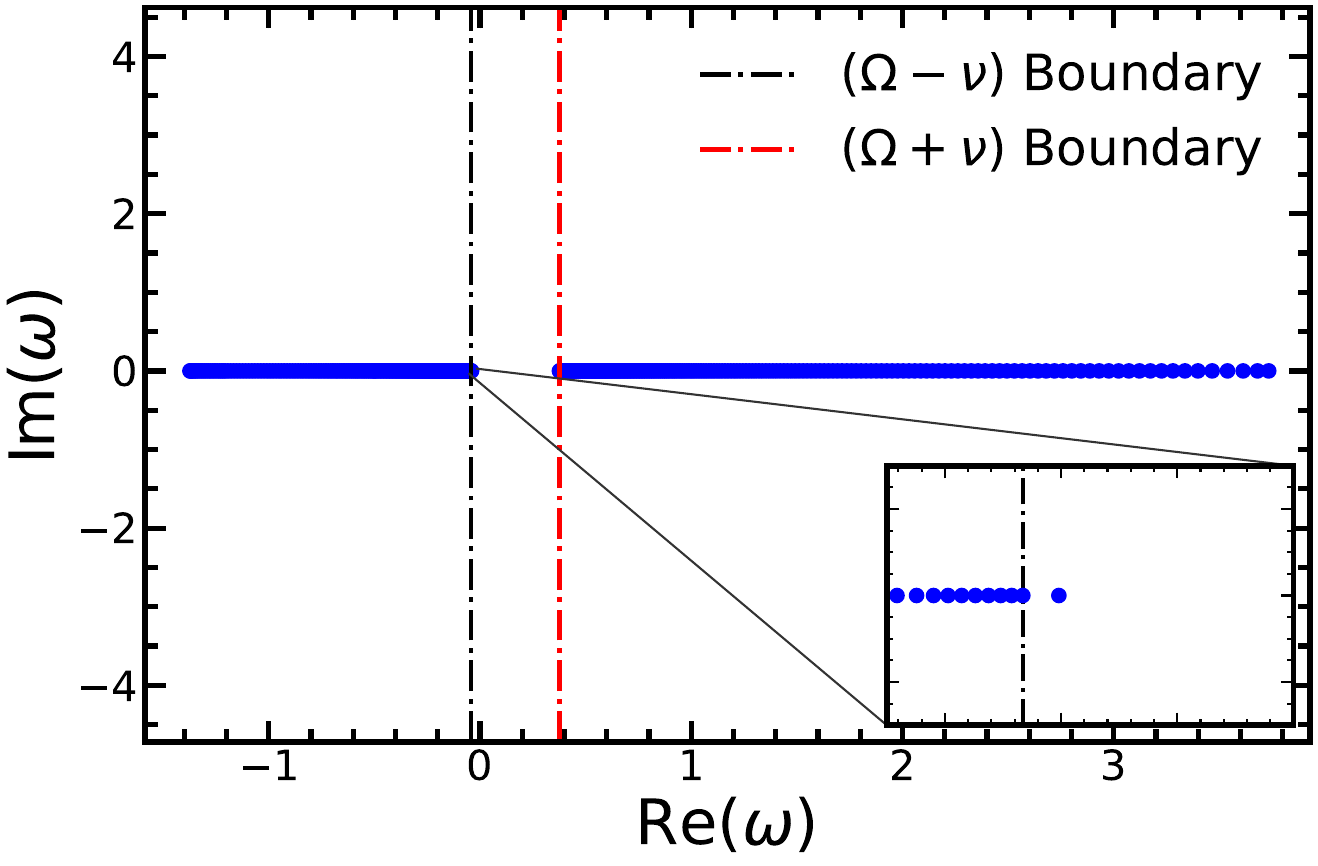}
\hspace{.65cm}
\includegraphics[width=.6\columnwidth]{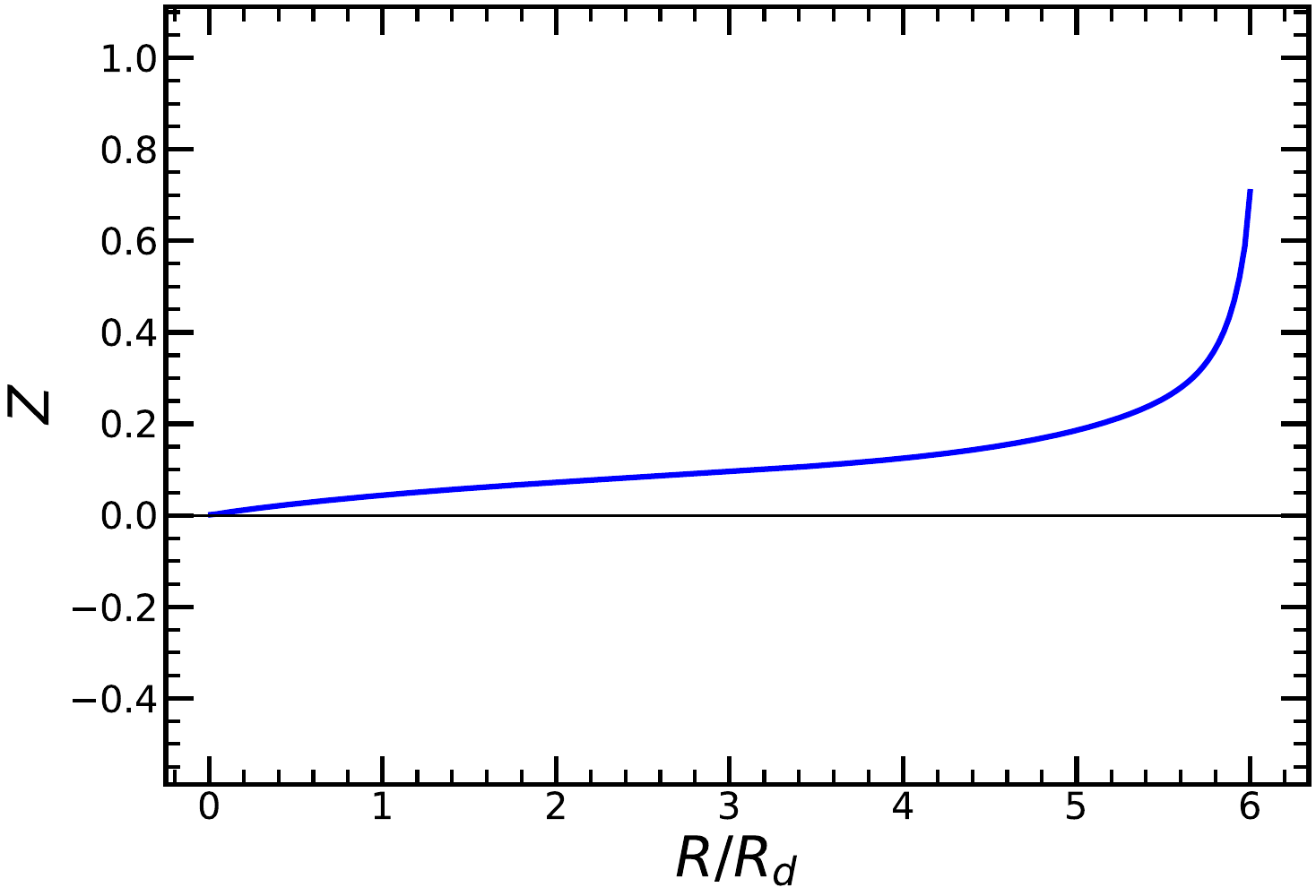}
\hspace{1cm}
\includegraphics[width=.55 \columnwidth]{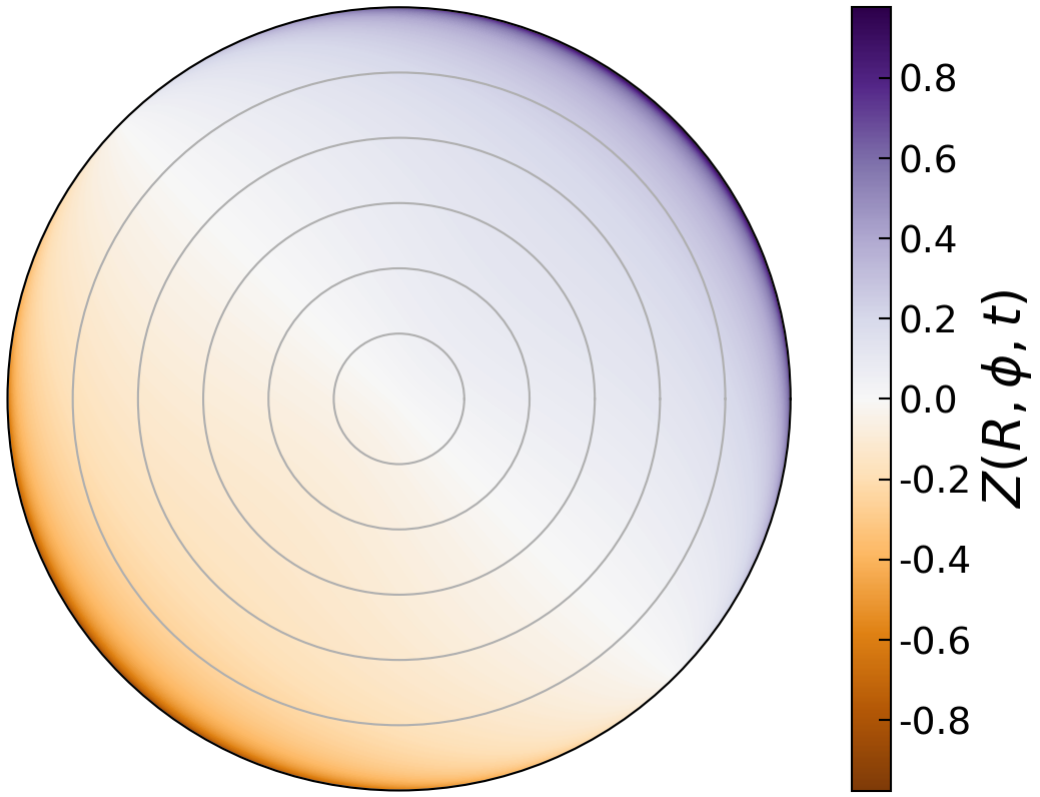}
 \caption{\textbf{Discrete mode:} Complete eigen spectrum and eigen function of  Model-I razor thin disk ($z_0=0$), in the absence of  vertical  pressure  and $\sigma_z$. The black and red dashed lines denote the lower and upper  boundary of the continuums $(\Omega -\nu)$ and $(\Omega+\nu)$ respectively.  The panel inside the left subplot shows the existence of an isolated eigen mode having eigen frequency value  Re$(\omega)=-0.04$  in the P-gap. The eigen function corresponding to that eigen mode resembles the long-range warp is shown in the middle subplot. The extreme right subplot shows the face-on map  of isolated eigen mode. The color bar next to the subplot shows the  vertical displacement $Z(R,\phi, t)$ in the unit of $R_d$. Concentric circles on  the right subplot indicate an increment of radius in the unit of $R_d$.}
\label{fig:Spectrum_1}
\end{figure*}

\begin{figure*}
\centering
\includegraphics[width=1.8\columnwidth]{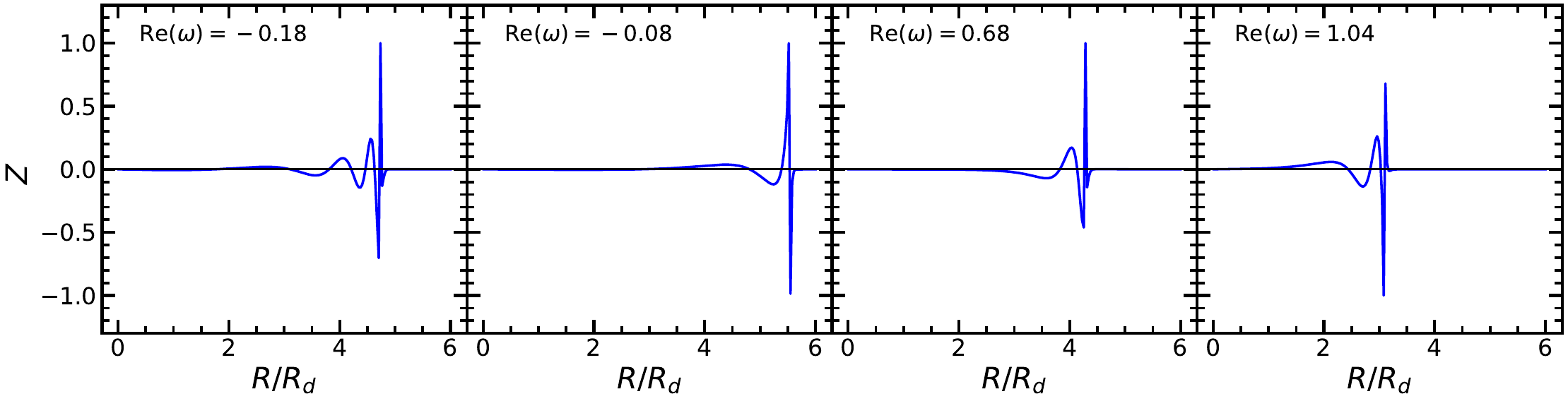}
    \caption{\textbf{Continuum mode shape:} Typical behavior of  eigen vectors corresponding to eigen modes lying in  $(\Omega-\nu )$  and $(\Omega+\nu )$ continuum  of  Model-I razor thin disk ($z_0=0$), in the absence of vertical pressure. The subplots of eigen functions are labeled with their corresponding  eigen frequencies. The negative eigen frequencies Re$(\omega)$ correspond to $(\Omega-\nu )$ continuum and the positive Re$(\omega)$ corresponds to $(\Omega+\nu )$ continuum.} 
    \label{fig:Stable_EF}
\end{figure*} 

\section{Numerical method and Input Parameters } \label{sec3}

\subsection{Numerical method for solving the equations}

To make numerical calculations convenient, we consider a system of units where $G = M_d = R_d = 1$. Throughout the analysis, all the frequencies and lengths are normalized in the unit of $\sqrt{GM_d/R_d^3}$ and $R_d$ respectively. We treat $(z-z')$ in the integral equation of $B_0(R,R')$ and $B_m(R,R')$ as a softening parameter to make integrals regular at $R=R'$. It is similar to the case of softening parameters used in \cite{1988MNRAS.234..873S} to make integral regular at $R=R'$  which avoids the divergence of numerical integration. The idea of softened- gravity was first introduced by \cite{1971Ap&SS..14...73M} and is used extensively in the numerical studies of disc dynamics. They replaced $1/R$ Keplerian potential by $1/(R^2+b^2)^{1/2}$ (where $b$ is the softening length) to avoid numerical singularities at $R = 0$. We keep the softening parameter equal to inter-ring spacing, which gives us a satisfactory result for the mode shape.

The equation (\ref{eqn:Int eqn})  can be solved by recasting it into a matrix-eigenvalue problem in a compact notation equation (\ref{eqn:QEP}) for the $m^{th}$ order bending mode. In the present paper, we focus only on the $m = 1 $ bending mode representing warps in disc galaxies. 

To find the solution of the QEP (\ref{eqn:QEP}), we treat the galactic disc as having a finite radius with a system of $N$ uniformly spaced concentric rings. In our analysis, the number of equally spaced concentric rings of the disc is considered as matrix dimensions. In a previous study of \cite{2008MNRAS.386L.101S} it is shown that the value of ground state discrete eigen frequency is not much affected by the  orders of matrix dimensions above $N=150$. In the present paper, we restrict the numbers of differentially rotating concentric rings of the disc or matrix dimensions  to $N=200$ only. The standard way to solve  QEP (\ref{eqn:QEP}) for the $m = 1$ mode is by reducing it to a generalized eigenvalue problem (GEP) of the form $Ax = \omega Bx$ by linearizing it into a $2N$ dimensional eigensystem (i.e., twice the matrix dimension $N$). We solve the linearized QEP numerically by using the standard technique for diagonalization. For further details about solving QEP, the readers are referred to \citep{hiki00q,saha2006generation}.  The Python module \textit{`scipy.linalg'} is used to solve the QEP equation and to find out all the eigen values and eigen functions. On solving the $N$-dimensional QEP (\ref{eqn:QEP}), we obtain $2N$ numbers of eigenvalues. The natures of eigen modes obtained from QEP (\ref{eqn:QEP})  for Model-I, Model-II, and Model-III, in the absence and in the presence of  vertical pressure gradient force are discussed in Section \ref{sec4}.

\subsection{Input Parameters}
In order to explore the bending instabilities in the disc,  we calculate the eigen modes considering  three sets of models that differ in the dark matter halo mass. At the same time, we consider two different vertical scale heights of the disc for each of the three models.  In Table.\ref{table:1} we give the disc and dark matter halo input parameters for the three sets of models required for numerical calculation in this paper. These input parameters of the disc and dark matter halo are arbitrary in order to first investigate the nature of eigen spectra and eigen modes. Later, we apply our analysis to three models of realistic Milky Way like  galaxy given in Table.\ref{table:MW}  from recent literature; Milky Way$_1$ \citep{2011MNRAS.414.2446M}, Milky Way$_2$ \citep{2013ApJ...779..115B}, and Milky Way$_3$ \citep{1998A&A...330..953M}. For dark matter halo, we use the parameters of  \cite{2010ApJ...712..260K} and explore the growth of  bending mode instabilities and wavelength in Section \ref{sec6}.

 \begin{table}  
\caption{The input parameters of the disc and dark matter halo of Model-I, Model-II, and Model-III are given. The system of units chosen has $M_{d}=R_{d}=1$. The vertical scale height is in the unit of $R_d$.  Time $t$ is normalized in the unit of $\tau = 2\pi/\omega$. } 
\label{table:1}
\centering
\begin{tabular}{c c c c c }      
\toprule
\toprule
Parameters & MODEL-I  & MODEL-II  & MODEL-III\\

    &       &       &           \\
\toprule
$z_0$	&	0.1	   &	0.1 &   0.1 \\
        &	 0.2	&	0.2 &   0.2 \\
\midrule
$M_h/M_d $	&	3	&	10  &	15 	\\
\midrule
$R_c/R_d$	&	3	&	3	&	3 \\
\midrule
$q$	    &	0.8	&	0.8 & 0.8 \\
\midrule
$t$	&	1/8 	&	1/8	&	1/8\\
\midrule
$R_t$	&	5 	&	5	&	5\\
\midrule
$R_o$	&	6 	&	6	&	6 \\

\bottomrule
\end{tabular}
\end{table}

 \begin{table}  
\caption{The input parameters of  three  models of Milky Way Galaxy and dark matter halo  from the available literature of various authors.} 
\label{table:MW}
\centering
\begin{tabular}{ c c c c}      
\toprule
\toprule
Parameters & Milky Way$_1$ & Milky Way$_2$ & Milky Way$_3$\\

         &    (PJ McMillan )    &  (Bovy \& Rix )    &  (Mera et al.)     \\
\toprule
$z_{_0}$  & 0.3~kpc&  0.4~kpc & 0.32~kpc\\

\midrule
$R_d$	& 3.29~kpc &2.15~kpc & 3.2 ~kpc	\\
\midrule
$M_d $ 	&  $6.6\times 10^{10}$~M$_{\odot}$ & $4.6 \times 10^{10}$~M$_{\odot}$ &  $4.6 \times 10^{10}$~M$_{\odot}$	\\
\midrule
$M_h $	&$1.59 \times 10^{11}$~M$_{\odot}$ &$1.47 \times 10^{11}$~M$_{\odot}$ & $5.6 \times 10^{10}$~M$_{\odot}$	\\
\midrule
$R_c$	 & 12~kpc &~12 kpc & 5~kpc\\
\midrule
$q$	     & 0.9&0.9 &	0.8 \\
\midrule
$V_0$	    & 239~km s$^{-1}$ & 230~km s$^{-1}$ &	220~km s$^{-1}$ \\

\bottomrule
\end{tabular}
\end{table}

\section{Numerical result and analysis} 
\label{sec4}

\subsection{Absence of vertical pressure force}
In order to analyze the $m = 1$ eigen modes in the absence of vertical pressure gradient, we assume that the disc is razor-thin. We solve the QEP (\ref{eqn:QEP}) for Model-I parameters considering only the radial surface density profile of the disc given in equation (\ref{eqn:disc_density}) and obtain the eigen values  and corresponding eigen vectors. In the left panel of Fig.\ref{fig:Spectrum_1}, we show the complete eigen spectrum for Model-I. From the eigen spectrum, we see that all the eigen modes are real i.e., all eigenmodes are stable. The eigen modes are stabilized by the gravitational restoring force due to the self-gravitating disc and dark matter halo (see the WKB dispersion in a later section).

\begin{figure*}
\includegraphics[width=1.6\columnwidth]{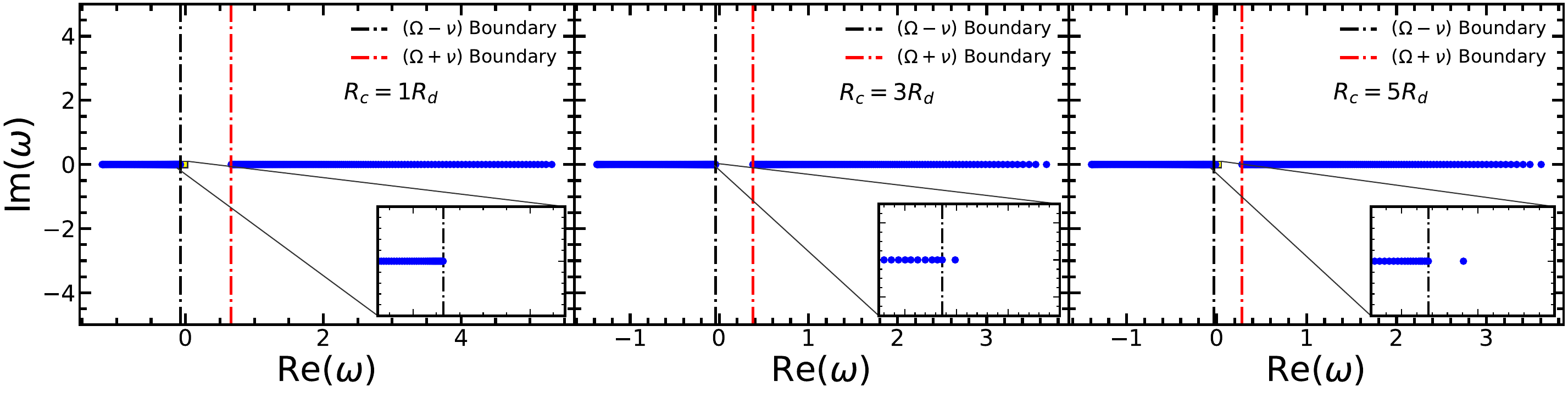}
\\
\hspace{3.5cm}
\includegraphics[width=1.2\columnwidth]{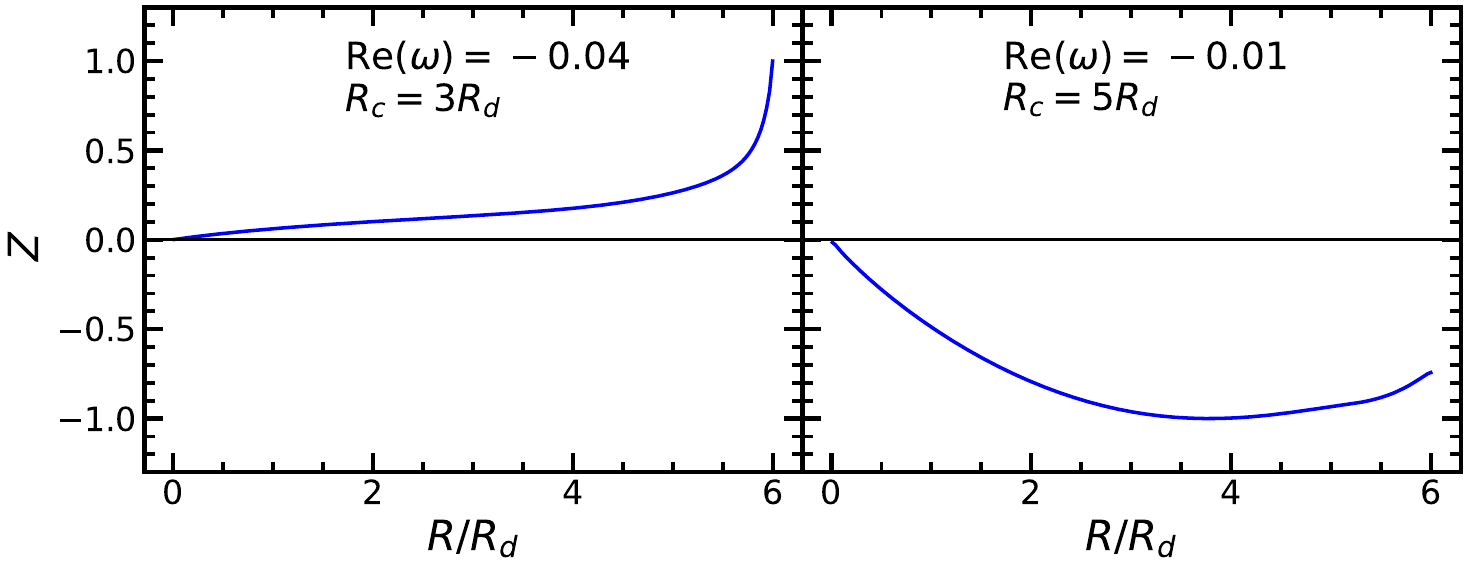}
\caption{ \textbf{Discrete mode at various $R_c$:} Eigen spectrum of Model-I razor-thin disc, showing the existence of discrete mode at three values of dark matter halo core radii $R_{c}= 1R_d$, $3R_d$, and $5 R_d$. The black and red dashed lines denote the lower boundary and upper of the continuum $(\Omega -\nu)$ and $(\Omega+\nu)$ respectively.  The inset zoom-in panels shows the discrete eigen mode lying in the P-gap between  the continuum.  For $R_c =1$, no discrete eigen mode exists in the P-gap. The bottom panel shows the shape of discrete modes for the halo core radii 3 and 5. }
\label{fig:E_SRc}
\end{figure*}

The eigen spectrum comprises two continuous branches of eigen modes with eigen frequencies $(\Omega-\nu)$ and $(\Omega+\nu)$, with a gap of length $2\nu$ between them. The right continuum is known as the fast mode $(\Omega+\nu)$, and the left continuum $(\Omega-\nu)$ is known as the slow mode \citep{2001AJ....121.1776T}. Except for a discrete mode, the two continua  contain all of the eigenmodes. The gap between these two continua is known as the principal gap (P-gap). The fast modes have positive eigen frequencies and are   prograde in the disc. The slow modes may withstand the differential shear in the disc for a longer period of time than the fast modes. 
   
The zoom-in inset subplot of the left panel shown in Fig.\ref{fig:Spectrum_1}  demonstrates the existence of a distinct eigen mode lying in the P-gap isolated from the continuum area. The mode corresponding to this point is a stable and stationary long-lived eigen mode and is immune to damping mechanisms such as wave-particle interaction known as Landau damping. This eigen mode is neither growing nor decaying and describes the global behavior of the integral shaped bending mode of the self-gravitating disc. This eigen mode behaves like a stable independent harmonic oscillator, vibrating at a frequency $\omega_R = -0.04$ with the lowest number of nodes (zero nodes in this case) along the radial axis. In the middle panel, we show the shape of the eigen vector corresponding to the discrete eigen mode. The face on map of the vertical  displacement $Z(R,\phi,t) = h(R)  \cos(\omega t-\phi)$ (real part of the equation (\ref{eqn:Z})) at $t = \tau/8 $ is shown in the right panel of Fig.\ref{fig:Spectrum_1} that replicates the large-scale, S-shaped warp in the disc. Here $\tau = 2\pi/\omega$ is the characteristic time scale of ground state eigen frequency of global mode in the P-gap. By ground state, we mean the mode having the lowest real part of the eigen frequency $\omega$ in the spectrum. The concentric rings are incremented along the radial axis from the center of the galactic disc with an amount of $1R_d$ disc scale length. The color bar right next to the plot indicates the amount of  displacement normal to the galactic plane in the unit of $R_d$. In Fig.\ref{fig:Stable_EF}, we show the typical behavior of eigen functions of a few selected eigen modes from the two continuum - these are akin to the singular van Kampen modes of oscillations in the disc \citep{2012arXiv1208.4338B}.
   
Although an isolated self-gravitating disc does not support a discrete bending mode $m=1$ until and unless the disc is truncated at the edge \citep{HunterToomre1969, Sellwood1996}, a smoothly tapered disc embedded in the oblate dark matter halo can support a discrete bending mode having the characteristics of large-scale warp in the disc. For such a system, the bending mode does not appear to be sensitive to details of the edge of the disc \citep{HunterToomre1969, Louis1992}. This result of finding a global discrete mode in the P-gap again confirms the fact that there exists at least one stable discrete normal mode \citep{Mathur1990, Weinberg1991,saha2006generation, 2008MNRAS.386L.101S} describing large-scale warp in the disc that exhibits stationary flapping oscillation in the $N$-ring system. The fact that the disc supports only stable eigen modes can be confirmed from the WKB dispersion relation of a small amplitude bending wave in the presence of self-gravitating force of the disc and dark matter halo force alone \citep{2008gady.book.....B}.
   
 \subsection{Effect of halo core radius on discrete eigen mode}
 
 Here, we explore the effect of halo core radius ($R_c$) on discrete eigenmode. The halo core radius governs how the halo mass is distributed in the disc and, ultimately, the frequency of vertical oscillation due to the halo. For a larger halo core radius, the circular velocity would rise slowly and a considerable fraction of the disc will reside inside the core. We solve the QEP (\ref{eqn:QEP}) for three configurations of the halo with core radii $R_c = 1 R_d, 3 R_d,$ and $5 R_d$ and obtain the corresponding eigen spectrum. In the top panel Fig.\ref{fig:E_SRc}, we show the complete eigen spectrum for these three cases. 
 
 The inset zoom-in plots shows the discrete eigen modes for each halo core radius. We find no discrete eigen mode in the P-gap for $R_c=1$, however for radii 3 and 5, a discrete eigen mode in the P-gap, distinct from the two continua is clearly evident.  The corresponding values of eigen frequencies for the value of core radius $R_c = 3 $ and  5 are $\omega_R = -0.04$ and $\omega_R = -0.01$ respectively. The value of discrete eigen frequency decreases with the increasing halo core radius. At a very smaller core radius,  the discrete modes merge into the $(\Omega-\nu )$ continuum. As a result, no discrete eigen modes are found in the eigen spectrum for halo core radius $R_c=1$.
 
 The shape of the discrete mode is highly sensitive to the halo core radius. The increasing core radius drastically changes the discrete mode shape. In the lower panel of Fig.\ref{fig:E_SRc}, we show the mode shape for two halo core radii $R_c= 3$ and 5. We obtain perfectly S-shaped bending for the core radius $3$. As the core radius decreases the shape or warp of the bending mode gets increasingly  curved at the edge of the disc while the inner region becomes flat.  At a large halo core radius, the discrete mode is unable to retain the S-shaped bending. In a nutshell, the halo core radius appears to play a fundamental role in supporting the long-scale S-shaped discrete mode in the disc. 
 
\begin{figure*}   
\includegraphics[width=.6\columnwidth]{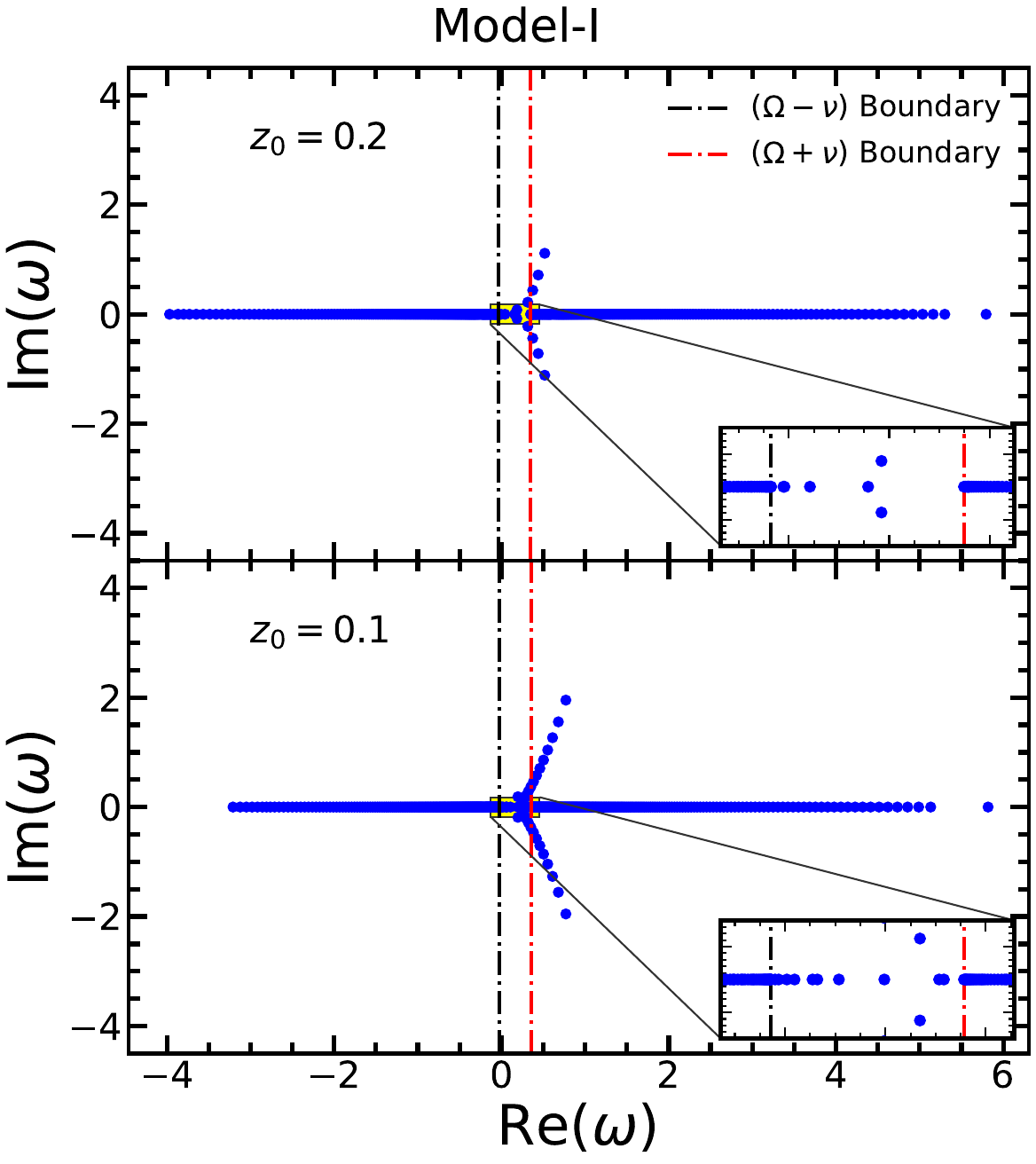}
   \hspace{.9cm}   
   \includegraphics[width=1\columnwidth]{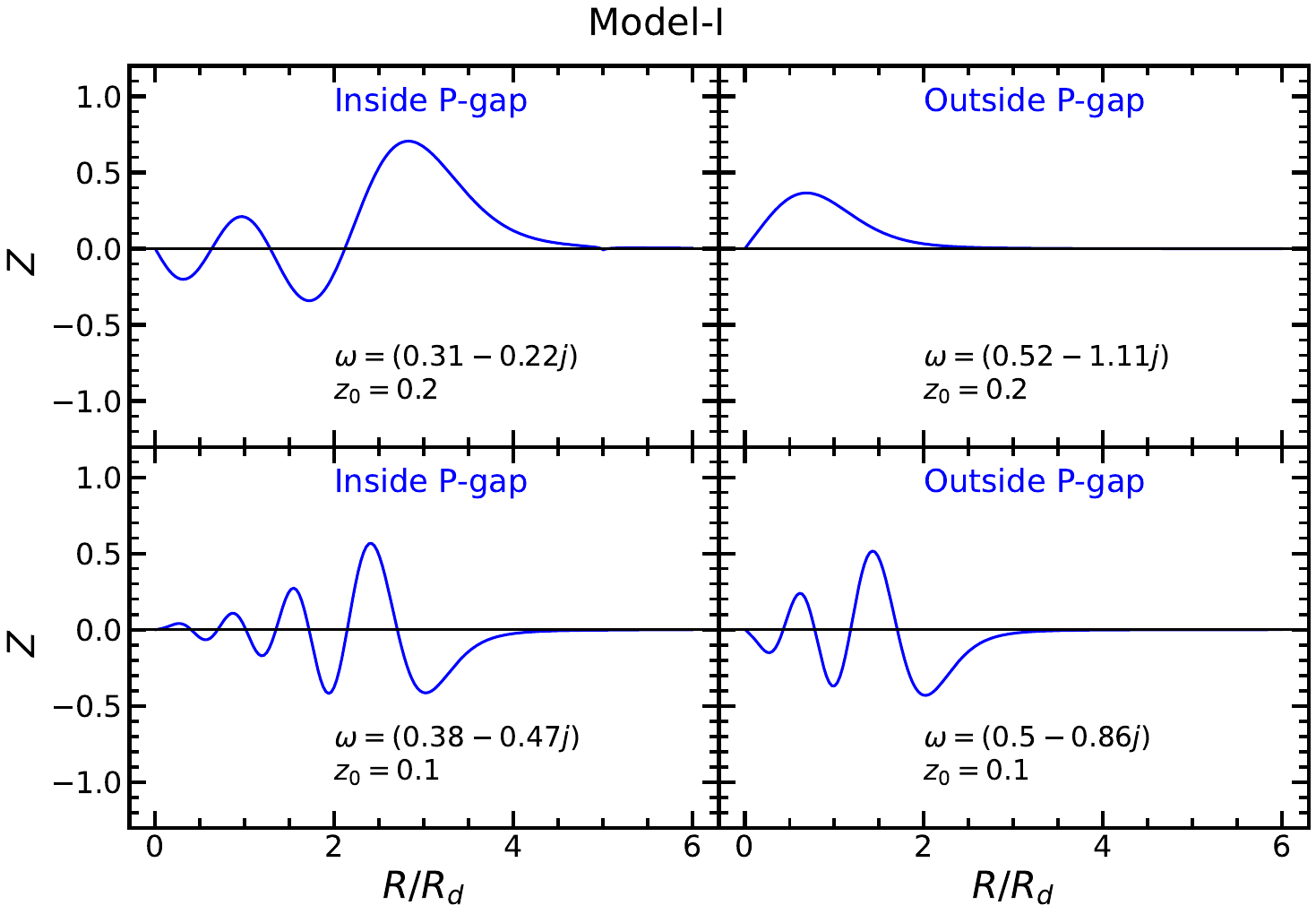}\\
   \includegraphics[width=.6\columnwidth]{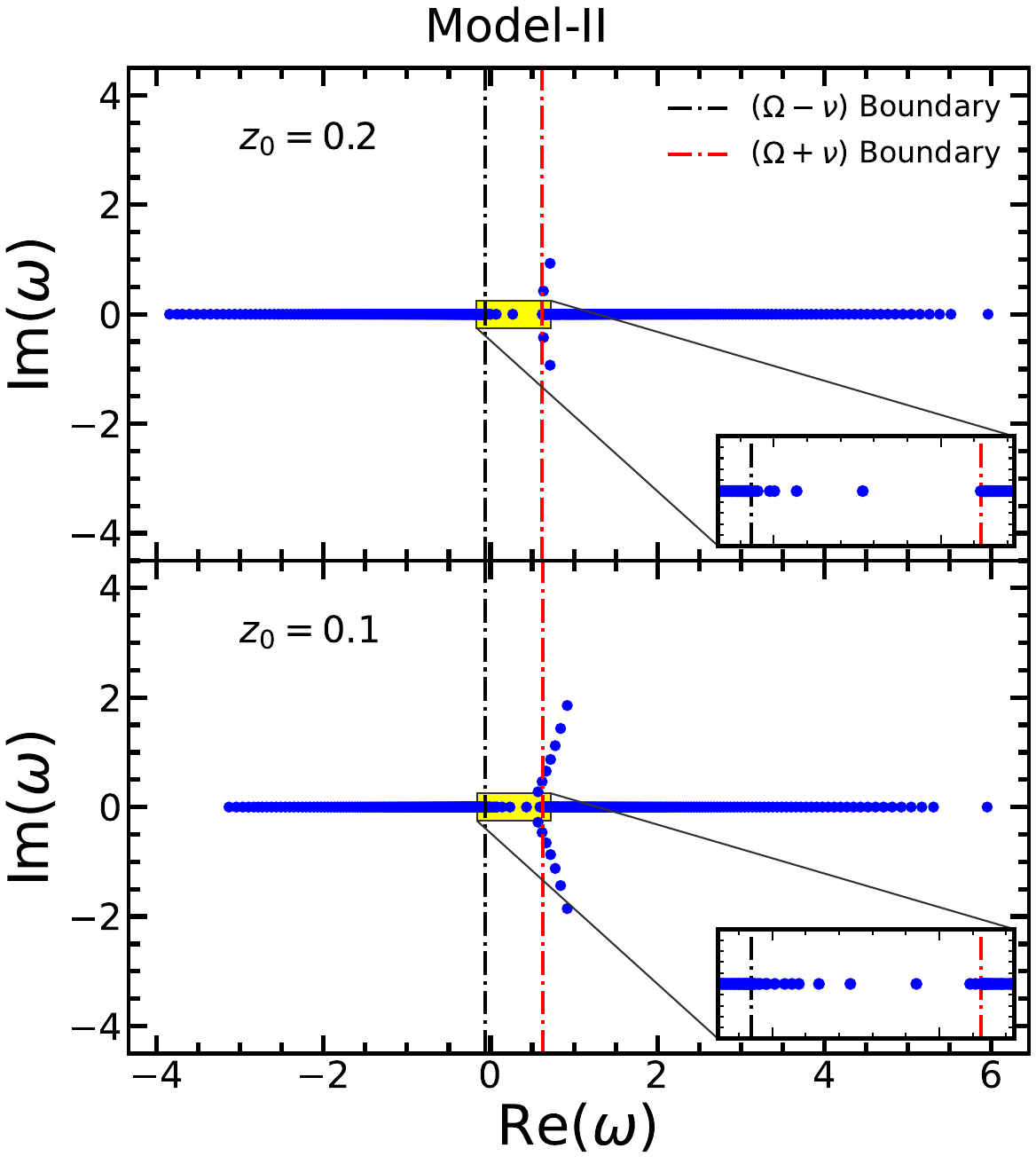}
    \hspace{.9cm} 
    \includegraphics[width=1\columnwidth]{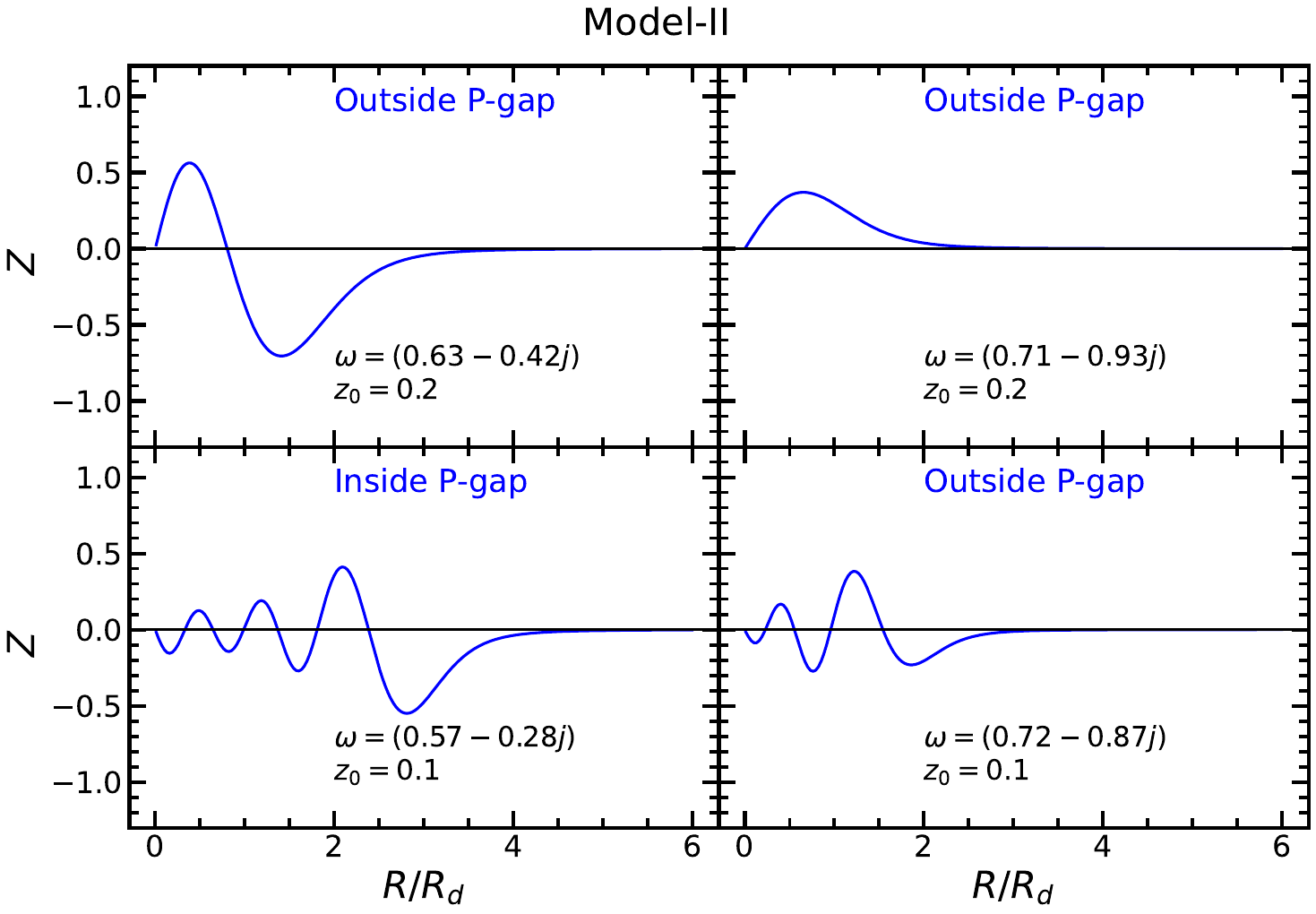}\\
   \includegraphics[width=.6\columnwidth]{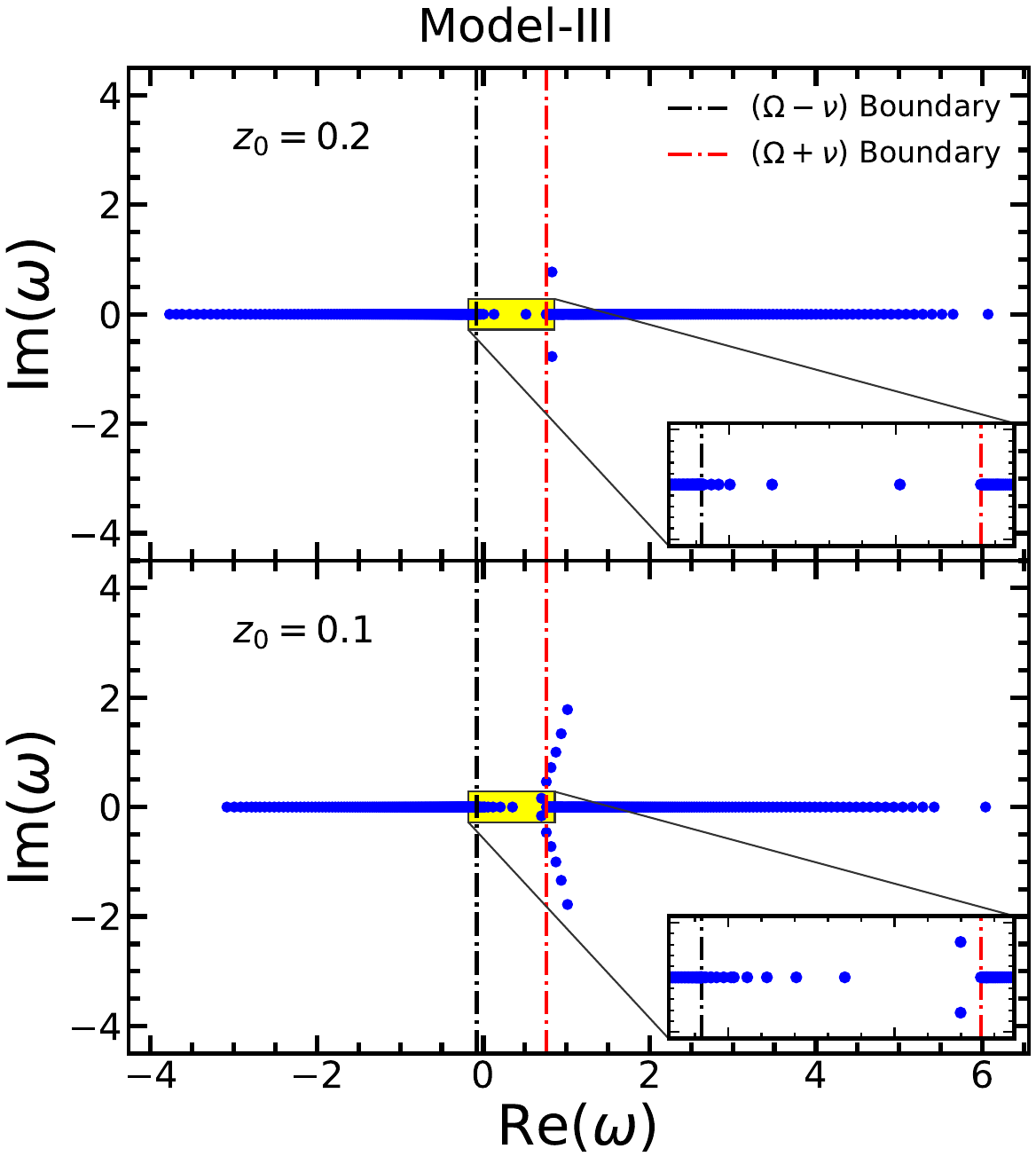}
   \hspace{.9cm} 
   \includegraphics[width=1\columnwidth]{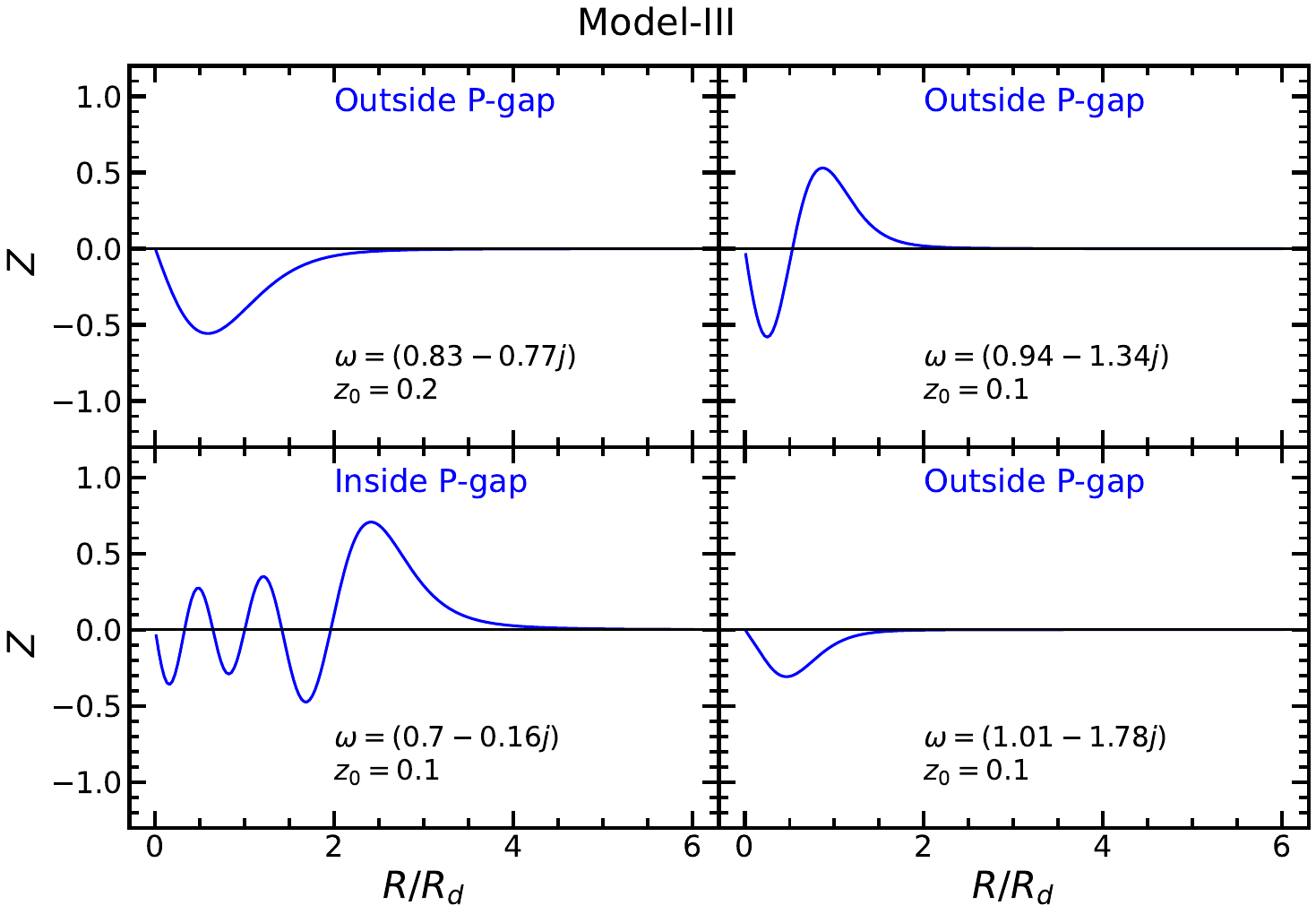}
    \caption{\textbf{Unstable eigen mode and shape:} Left panel: complete eigen spectrum for Model-I, Model-II, and Model-III, showing stable as well as unstable modes in the presence of vertical pressure. For each model, we show the eigen spectrums for $z_0=0.2$ (top panel) and $z_0 =0.1$ (bottom panel). The black and red dashed lines denote the lower and upper boundary of the continuum $(\Omega -\nu)$ and $(\Omega+\nu)$ respectively.  The inset panels show the stable and unstable eigen modes lying in the P-gap. Right panel: mode shape of a few selected unstable modes lying inside and outside the P-gap of each model. }
    \label{fig:Eigen Spectrum Model I - II}
\end{figure*}

\begin{figure*}
\centering   
    \includegraphics[width=1.6\columnwidth]{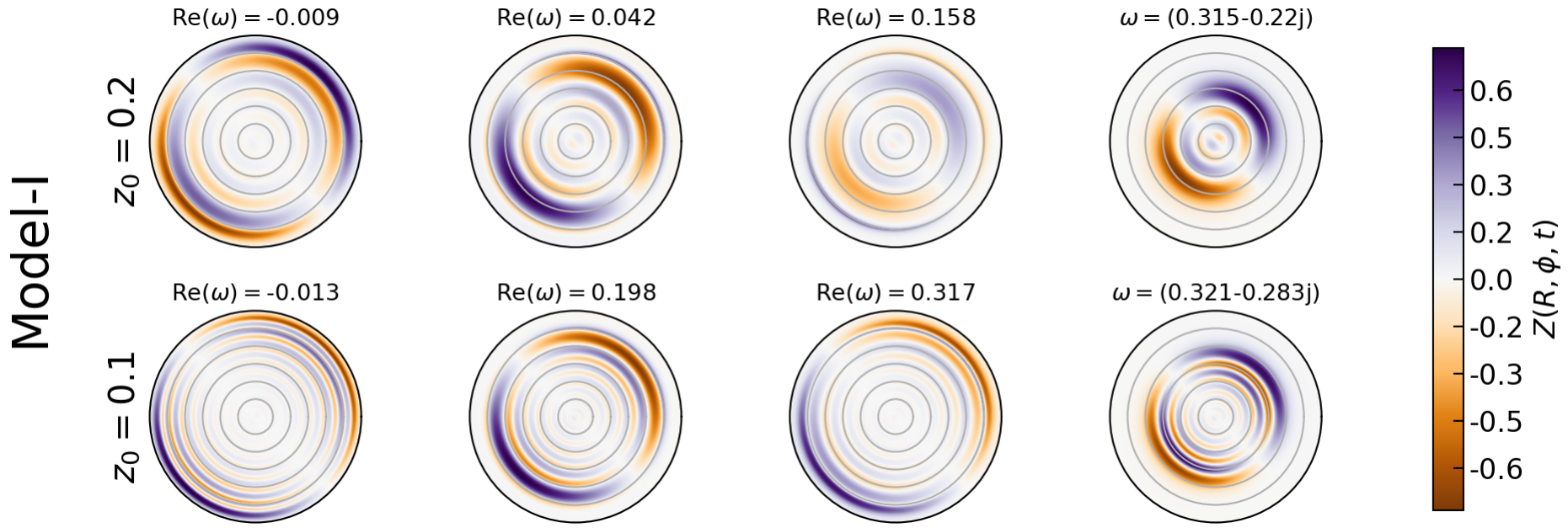}\\
    \vspace{.5cm}
    \includegraphics[width=1.6\columnwidth]{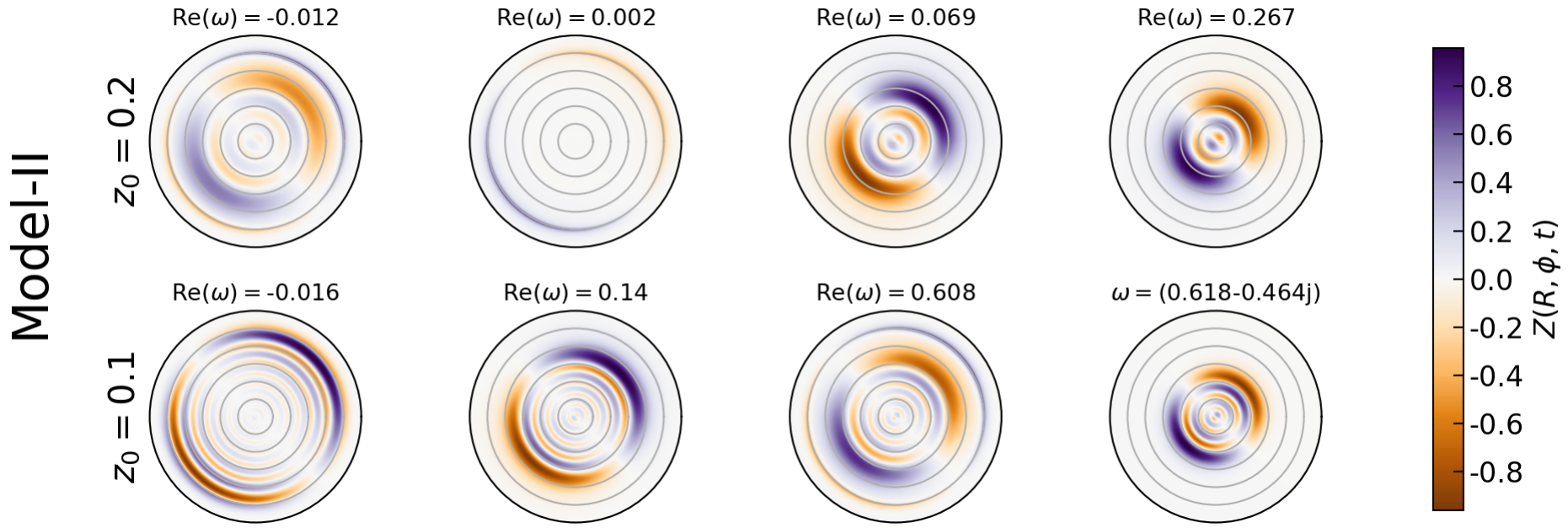}\\
    \vspace{.5cm}
    \includegraphics[width=1.6\columnwidth]{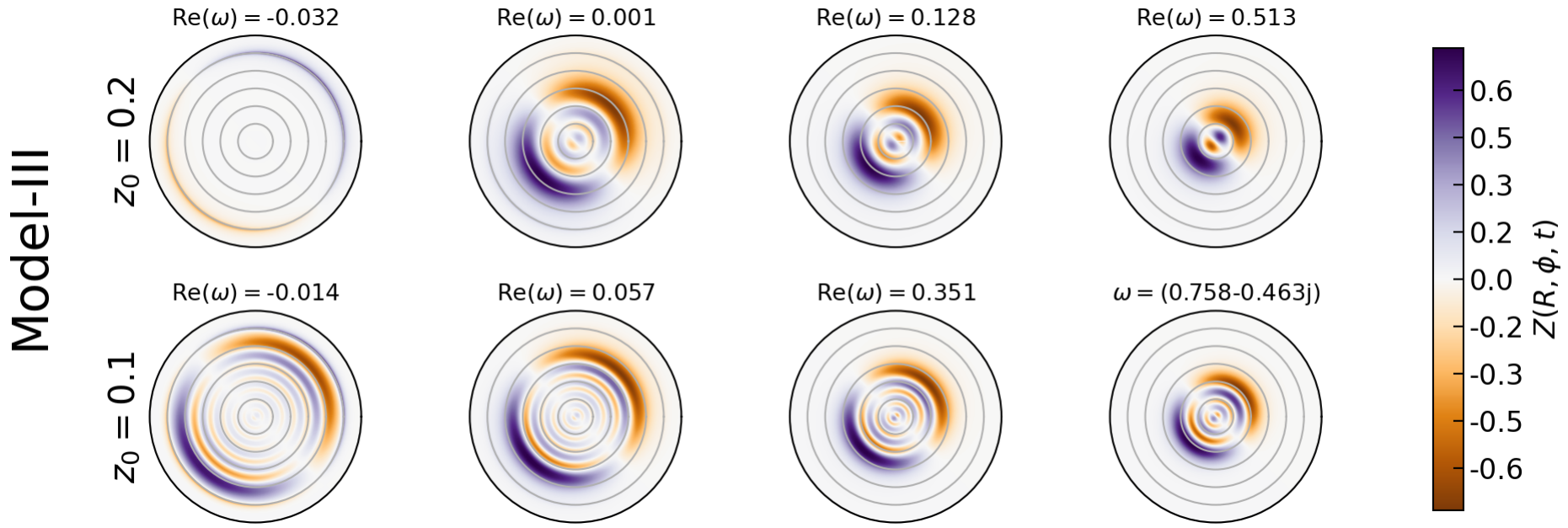}
    \caption{\textbf{Discrete modes shape:} Face-on maps of the vertical displacement $Z(R,\phi,t)$  of  eigen modes lying in the P-gap of all the three models. The eigen frequencies for each mode are  shown at the top of each subplot. Concentric circles on  subplots indicate increments of  $1R_d$ in radius. The color bars  right next to subplots indicate $Z(R,\phi,t)$ in the unit of $R_d$. The extreme right subplots show the unstable modes  in the P-pap (except for $z_0 =0.2$ in Model-II and III since no unstable modes exist in the P-gap in these models).}
    \label{fig:Map-I}
\end{figure*}
   
\section{Eigen modes in the presence of vertical pressure}\label{sec5}

In this section, we attempt to investigate the properties of eigen modes in more realistic models of disc galaxies by including the vertical pressure gradient force calculated self-consistently as shown in section~\ref{sec:verticalpressure}. The radial profile of the vertical velocity dispersion $\sigma_z (R)$ is shown in Fig.\ref{fig:Density-Dispersion}. We explore all three models described in Table\ref{table:1}. The rotation curves for each halo to disc mass ratio and disc scale heights of all three models producing reasonably flat rotation curves are shown in Fig.\ref{fig:R_C}. Each of these models is primarily characterized by their different halo-to-disc mass ratio ($M_h/M_d$). In the following, we describe our findings for each model. 

\begin{table*}
\caption{\textbf{Time period and numbers of discrete stable and unstable modes:} The highest rotation period $T_{R=6}$ of the disc at radial distance $R=6 R_d$, time periods of lowest  mode frequency  $\left(T_{low}=2\pi/|\omega_{low}|\right)$  and numbers of stable, unstable modes inside P-gap and unstable modes outside the P-gap of the three models.  The unit of time-scale for the fiducial disc is $\sqrt{R_d^3/GM_d}=15.58$ Myr.} 
\label{table: time period}
\centering
\begin{tabular}{c c c c c c c c}     
\toprule\toprule
   Model  &$z_0 $ &  $T_{R=6}$ & No. of stable modes & No. of unstable modes & No. of unstable modes &$\omega_{low}$ & $T_{low}$\\
            &       &   (Gyr)      &  inside  P-gap    &   inside P-gap  &  outside P-gap  &         & ~(Gyr) \\
          
\hline
 Model- I & 0.1   & 0.489      & 10 &   5 &   8 &-0.013     & 7.53\\
          & 0.2   & 0.469      & 4  &    2  &  3 &-0.009     & 10.87\\
\midrule
Model-II & 0.1    & 0.273      & 15 &   2     & 5 &-0.016     & 6.11\\
         & 0.2    & 0.269      & 6 &    0     & 2 &-0.012     & 8.15\\
\midrule
Model- III & 0.1  & 0.225      & 12 &   2      & 4& -0.014    & 6.99\\
           & 0.2  & 0.222      & 7 &    0     & 1 &-0.032     & 3.05 \\
 \bottomrule
\end{tabular}
\end{table*}
\subsection{Model-I}

On the top left panel of Fig.\ref{fig:Eigen Spectrum Model I - II}, we show the eigen spectrum of the disc having the vertical disc scale heights $z_0= 0.2$ (top panel) and  $z_0= 0.1$ (bottom panel). The inset plots highlight the existence of discrete stable eigen modes in the P-gap for discs with both vertical scale heights. The complex eigenvalues are distributed in a wedge-like fashion on the Argand diagram. In either case, there exist both stable and unstable eigenmodes lying in the P-gap with a slightly lesser number of unstable modes  in $z_0=0.2$ in comparison to that $z_0 =0.1$ (see Table.\ref{table: time period}). Some of these unstable eigen modes are shown on the top right panel of Fig.~\ref{fig:Eigen Spectrum Model I - II}. The stable discrete modes have a shape as shown in Fig.~\ref{fig:Map-I} and have low-frequency oscillations \citep{2008MNRAS.386L.101S}. 
The top panel of Fig.\ref{fig:Map-I} displays the face-on map of a few discrete eigen modes found in the P-gap of Model-I. The panels are labeled with the value of eigen mode frequencies $\omega$. The extreme left panels show the slowest discrete mode in the P-gap. The extreme right panels show the unstable modes present in the P-gap. The color bar shows the scale of displacement $Z(R,\phi,t)$ along the vertical $z-$ axis in the unit of $R_d$. The lowest frequency mode for model-I with $z_0 = 0.2$ and $z_0 = 0.1$ have the time-period $T_{low} =10.87$~Gyr and 7.53~Gyr respectively,  considering fiducial disc model parameters (see Sec.\ref{sec6} for a further discussion). The unit of time-scale for the fiducial disc is $\sqrt{R_d^3/GM_d}=15.58$ Myr (see Table \ref{table: time period}). As the lowest frequency mode has a larger time period than the disc rotation period such modes are retrograde in the disc.  

\begin{figure}
    \centering
    \includegraphics[width=.9\columnwidth]{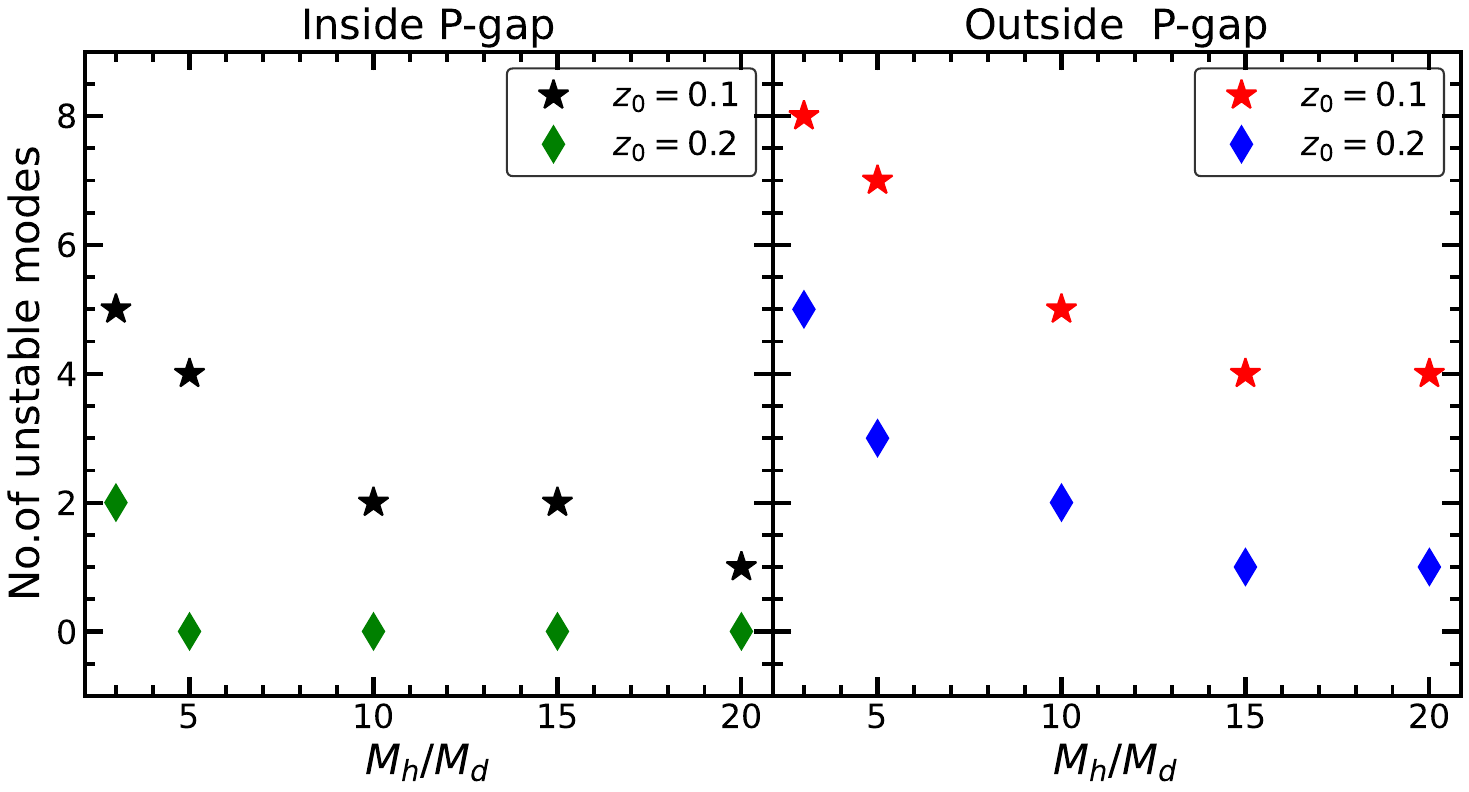}
    \caption{\textbf{No. of Unstable Modes:} The number of unstable modes inside and outside the P-gap for the halo masses $M_h/M_d= 3, 5, 10, 15, 20$ for $z_0 =0.1$ and 0.2. The plot shows the number of unstable modes decreases with increasing halo mass and thickness of the disc.}
    \label{fig: no.of unstable modes}
\end{figure}

The eigen modes with negative imaginary parts are of particular interest because they grow exponentially with time influencing overall disc structure. The unstable modes in the eigen spectrum arise due to the finite vertical velocity dispersion of the disc (as can be seen from the WKB relation given below). The eigenmodes with positive imaginary parts are damped and are absorbed by the disc particles in the form of random kinetic energy - which in turn would heat the disc and might affect the further instability of the disc through feedback \citep{NelsonTremaine1995}. None of these unstable modes resemble those in the continuum similar to the singular van Kampen modes (see Fig.~\ref{fig:Stable_EF}). We notice that in the case of $z_0=0.1$, the number of eigenmodes in the P-gap is higher as compared to the disc with $z_{0}=0.2$. Apart from the discrete  stable modes,  the vertical pressure excites a few unstable in the P-gap which are comparatively higher in number for $z_0=0.1$ (see Table.\ref{table: time period} and Fig.\ref{fig: no.of unstable modes}).  Coming to the unstable growing mode  we obtain  5 unstable modes for the $z_0=0.2$ case, whereas the disc with $z_0=0.1$ has 13 growing modes (see Table.\ref{table: time period}). As the disc thickens, the unstable modes start disappearing (see Fig.\ref{fig: no.of unstable modes}) confirming previous findings by \cite{Sellwood1996}. We estimate the wavenumber and wavelength of these unstable modes using Fourier transform \citep{2002nrca.book.....P} and discuss this aspect in greater detail in the light of WKB relation in Sec.~\ref{sec:WKB}.

\begin{figure*}
\centering
\includegraphics[width=1.6\columnwidth]{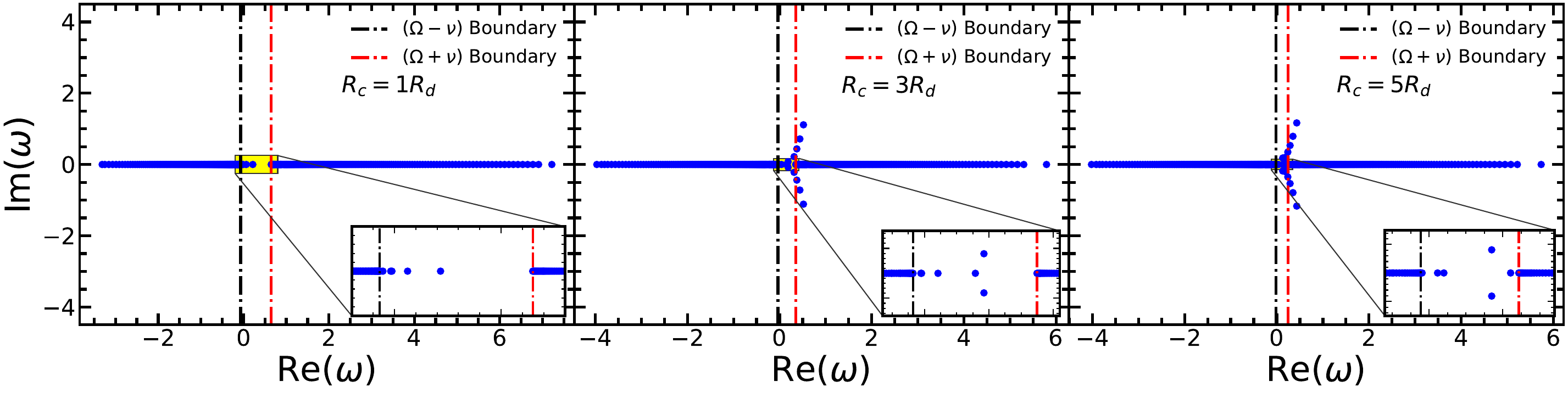}\\
\vspace{.5cm}
\includegraphics[width=0.8\columnwidth]{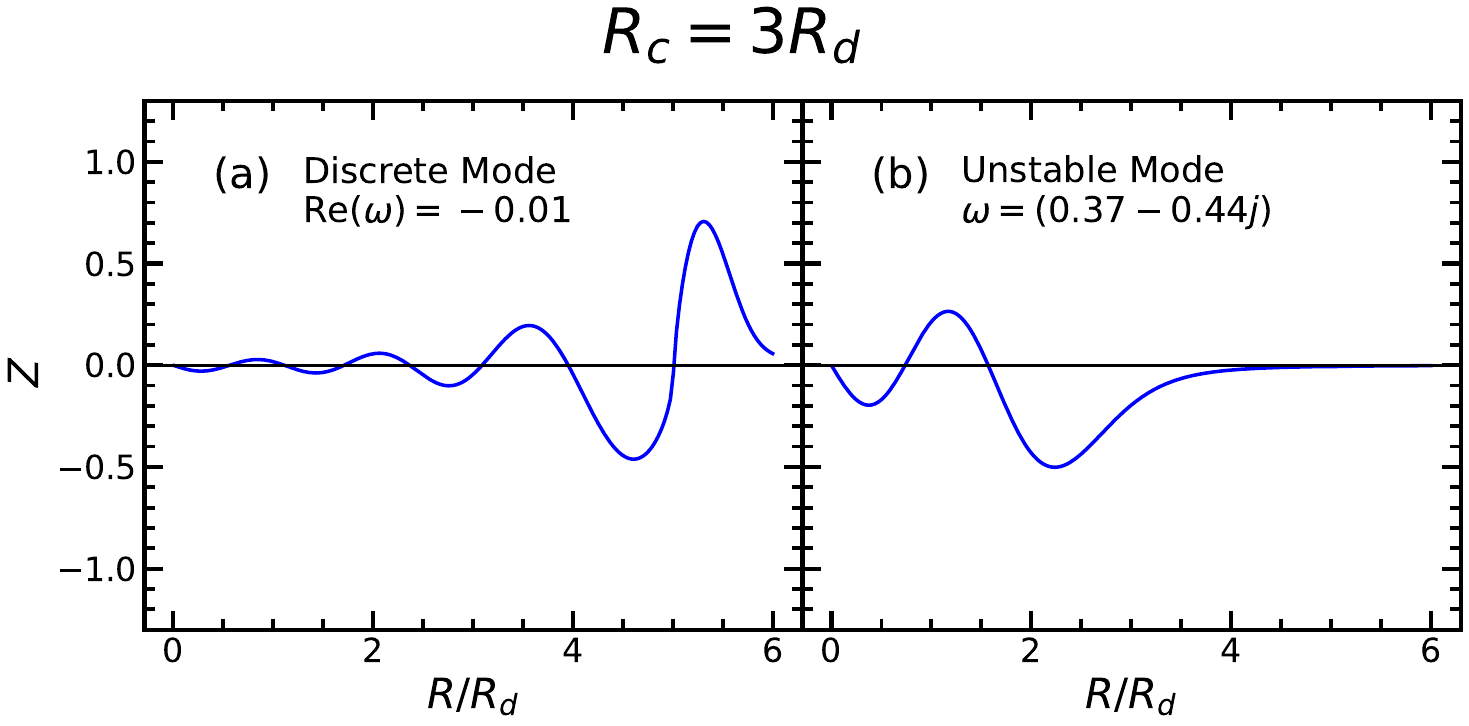}
\includegraphics[width=0.8\columnwidth]{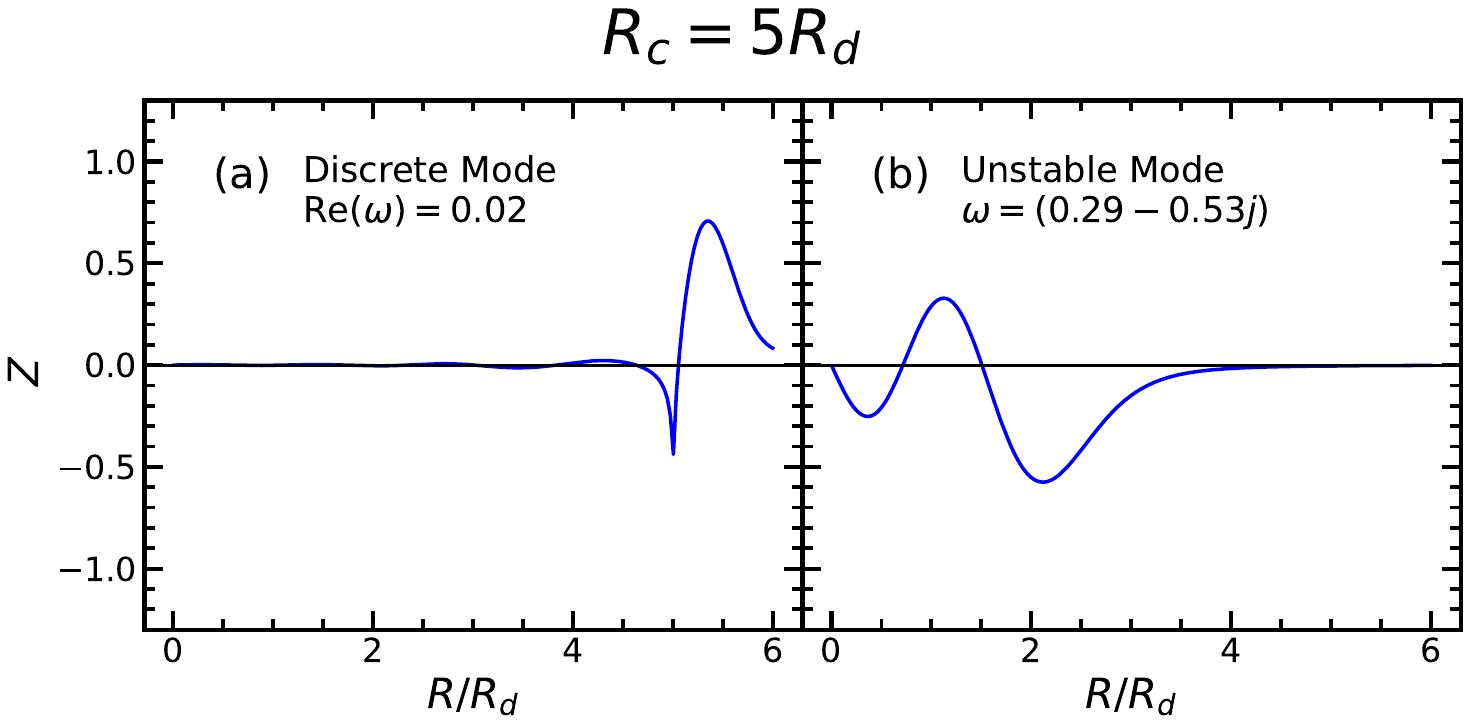}

\caption{\textbf{Discrete and unstable modes at various $R_c$:} Top panel:  Model-I ($z_0=0.2$) eigen spectrum  for three values of dark matter halo core radii $R_{c}= 1R_d$, $3R_d$, and $5 R_d$,   in the presence of vertical pressure. The black and red dashed lines denote the lower boundary and upper of the continuum $(\Omega -\nu)$ and $(\Omega+\nu)$ respectively.  The inset zoom-in panels show the stable and unstable discrete eigen modes lying in the P-gap. Bottom panel: the shape of  (a) discrete stable mode inside the  P-gap and  (b) unstable modes outside the P-gap for the core radii 3 and 5.}
\label{fig:PES_Rc}
\end{figure*}

\subsection{Model-II}
The Model-II is similar to Model-I but with a dark matter halo mass $M_h  =10 M_d$, in other words, the disc is more dominated by the dark halo compared to Model-I. The eigen spectrum of Model-II is shown in the middle left panel of Fig.\ref{fig:Eigen Spectrum Model I - II} for both the vertical scale heights $z_0 = 0.1$, $0.2$. The face-on maps of $Z(R,\phi,t)$ for a few selected discrete stable modes are shown in the middle panel of Fig.\ref{fig:Map-I}. These discrete modes have fewer radial nodes as compared to the eigen modes in the two continua. The typical behavior of discrete mode is similar to that of Model-I (see Fig.\ref{fig:Eigen Spectrum Model I - II}). Similar to Model-I, the lowest frequency discrete mode has a significantly larger time period (see Table. \ref{table: time period})  than the disc rotation period that avoids resonance in the disc. 

Although the trends of the eigen spectrum are similar to Model-I, the number of discrete stable modes in the P-gap and unstable modes are different. For the scale height $z_0 =0.2 $, only 2 unstable modes are obtained whereas, there are 7 unstable modes in the case of $z_0 =0.1$. It is worth noting  that the unstable mode in $z_0=0.1$ in the P-gap as well as outside the gap  is more than that of $z_0=0.2$ following the same trends as the previous model. The number of unstable modes in the P-gap as well outside the gap reduces in this model (see Table.\ref{table: time period} and Fig.\ref{fig: no.of unstable modes}).  Since the combination of disc self-gravity and enhanced restoring force due to the increased dark matter halo mass has a stabilizing effect, it is natural to expect a number of decreased unstable modes inside and outside the P-gap. In other words, these results just reaffirm that disc instabilities are, in general, suppressed by a massive dark matter halo \citep{ostriker1973numerical} but how the unstable modes disappear remains a question. Note that this is true only when the halo is non-responsive, else a live halo might lead to excitation of a bending mode \citep{NelsonTremaine1995,WeinbergBlitz2006} in the disc outskirts or bar growth \citep{Athanassoula2002,WeinbergKatz2007,saha2013spinning} in the central region of the galaxy.

\subsection{Model-III}
The dark matter halo mass ($M_h =15 M_d $) in Model-III is even higher as compared to Models-I and II. The eigen spectrum for this model is shown in the bottom left panel of Fig.\ref{fig:Eigen Spectrum Model I - II}. The subplots of the eigen spectrum are labeled for the different values of disc scale heights $z_0$. We obtain similar trends of the eigen spectrums as observed in Model-I and II but with different numbers of eigen modes. 

For $z_0=0.2$, only one  unstable  mode is obtained outside the P-gap. Being more massive, the dark matter halo is able to suppress the unstable modes even further. Whereas, the disc with $z_0=0.1$ still has 6  unstable modes: 4 outside the P-gap and 2  inside the P-gap. The typical shape of unstable modes is shown in the bottom right panel of Fig.\ref{fig:Eigen Spectrum Model I - II}. The number of stable oscillating discrete eigen modes is found to be decreasing for  the $z_0 = 0.1$ value in this model as compared to the previous two models. The face on maps of a few selected discrete modes are shown in the bottom panel of Fig.\ref{fig:Map-I}.

From the overall analysis, we see that for a particular disc scale height, low-mass halos support larger number of unstable modes (both inside and outside the P-gap) compared to their massive counterpart (See Fig.~\ref{fig: no.of unstable modes}). Thinner disc supports more unstable modes compared to the thicker discs. For example, for $z_{0}=0.1$, the disc still supports unstable modes for halo mass as high as $M_h/M_d =20$. For $z_{0}=0.2$ case, although unstable modes cease to exists inside the P-gap, a few of them still persist outside the P-gap. It is interesting to note that as we increase halo mass, the unstable modes in the disc first disappear from the P-gap and later from outside the P-gap. Based on several numerical works \cite{ostriker1973numerical, Efstathiouetal1982, YurinSpringel2015}, it is known that a stellar disc is stabilized by a massive dark matter halo but the process in which discs are stabilized remained obscure. Our simplified analysis based on analytical work, shed some light in this regard.

\subsection{Effect of dark matter halo core radius on instabilities}

For the sake of completeness, we calculate and present the eigen spectrum of  Model-I input parameters having disc scale height $z_0=0.2$ for three different dark matter halo core radii. The primary goal is to investigate the number of unstable modes that arise in the eigen spectrum with the increasing halo core radii. In Fig.\ref{fig:PES_Rc}, we show three eigen spectrums for three different halo core radii $1R_d$, $3R_d$, and $5R_d$.  We obtain no unstable modes for $R_c=1$ whereas, we obtain 5 and 7 unstable modes for $R_c = 3$ and 5 respectively. Note that when the disc thickness is reduced to $z_0=0.1$, we obtain a few unstable modes in the eigen spectrum for core radius $R_c=1$. The number of unstable modes in the eigen spectrum is found to be increasing with the increase of the halo core radius. In other words, the disc is susceptible to $m=1$ bending instability when embedded within a dark matter halo with a larger core radius. As the halo core radii increase, the dark matter halo becomes less concentrated in the inner region of the disc which in turn weakens the gravitational restoring force on the disc due to the halo. As a consequence, the disc is unable to counterbalance the destabilizing force due to the vertical pressure. We have verified our numerical results for Model-II and III and found similar trends as in Model-I. In the bottom panel of Fig.\ref{fig:PES_Rc}, we show the typical nature of the discrete and unstable mode shape in subplots (a) and (b) for the two  halo core radii $R_c=3$ and $5$ respectively.

\begin{figure*}
\centering
\includegraphics[width=1.8\columnwidth]{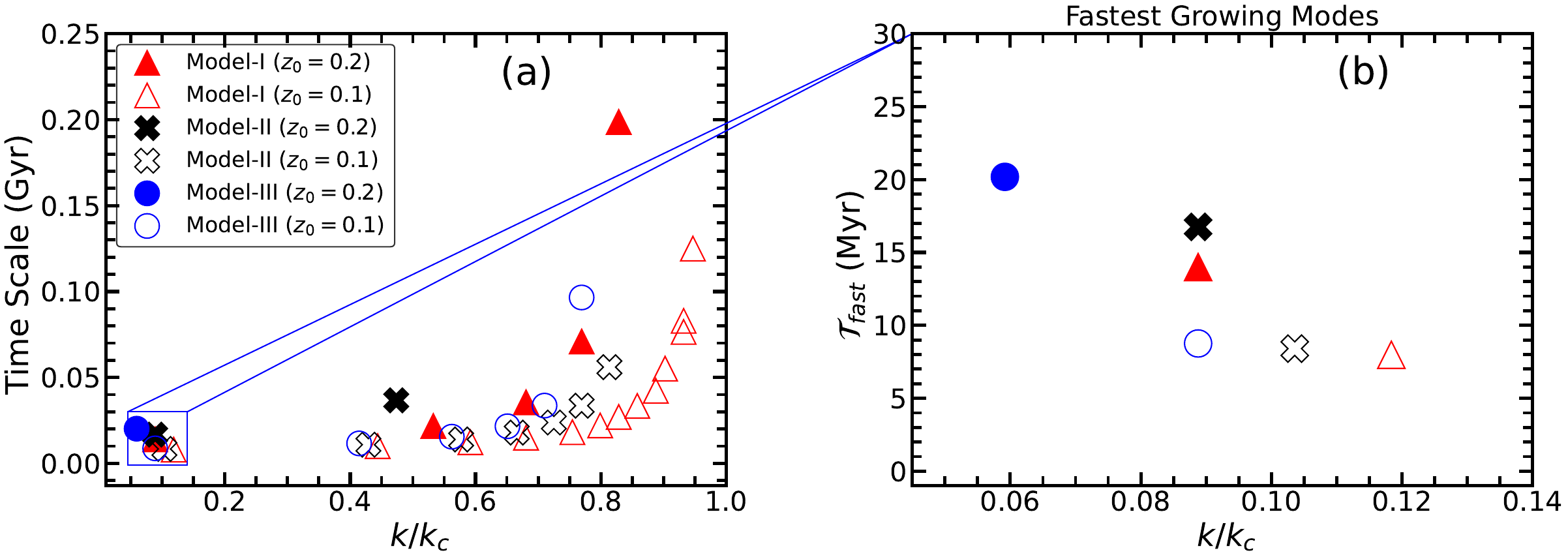}
    \caption{\textbf{Growth time scale and wavenumber of unstable modes:} Growth time scale of the unstable modes of the fiducial  Model-I, Model-II, Model-III. The three models have disc mass $M_d = 3 \times10^{10}$ M$_\odot$ and scale radius  $R_d = 3.2$ kpc. In panel (a) we  show the growth time scale in Gyr for all the unstable modes corresponding to its  dimensionless wavenumber $k/k_c$. In the right panel (b) we show the growth time scale ($\mathcal{T}_{fast}$) in Myr of the fastest growing modes for the three  models. Here the unit of wave number $k$ and critical wavenumber $k_c$ is kpc$^{-1}$.}
    \label{fig:timescale}
\end{figure*}

 \begin{figure*}
\centering
\includegraphics[width=1.6\columnwidth]{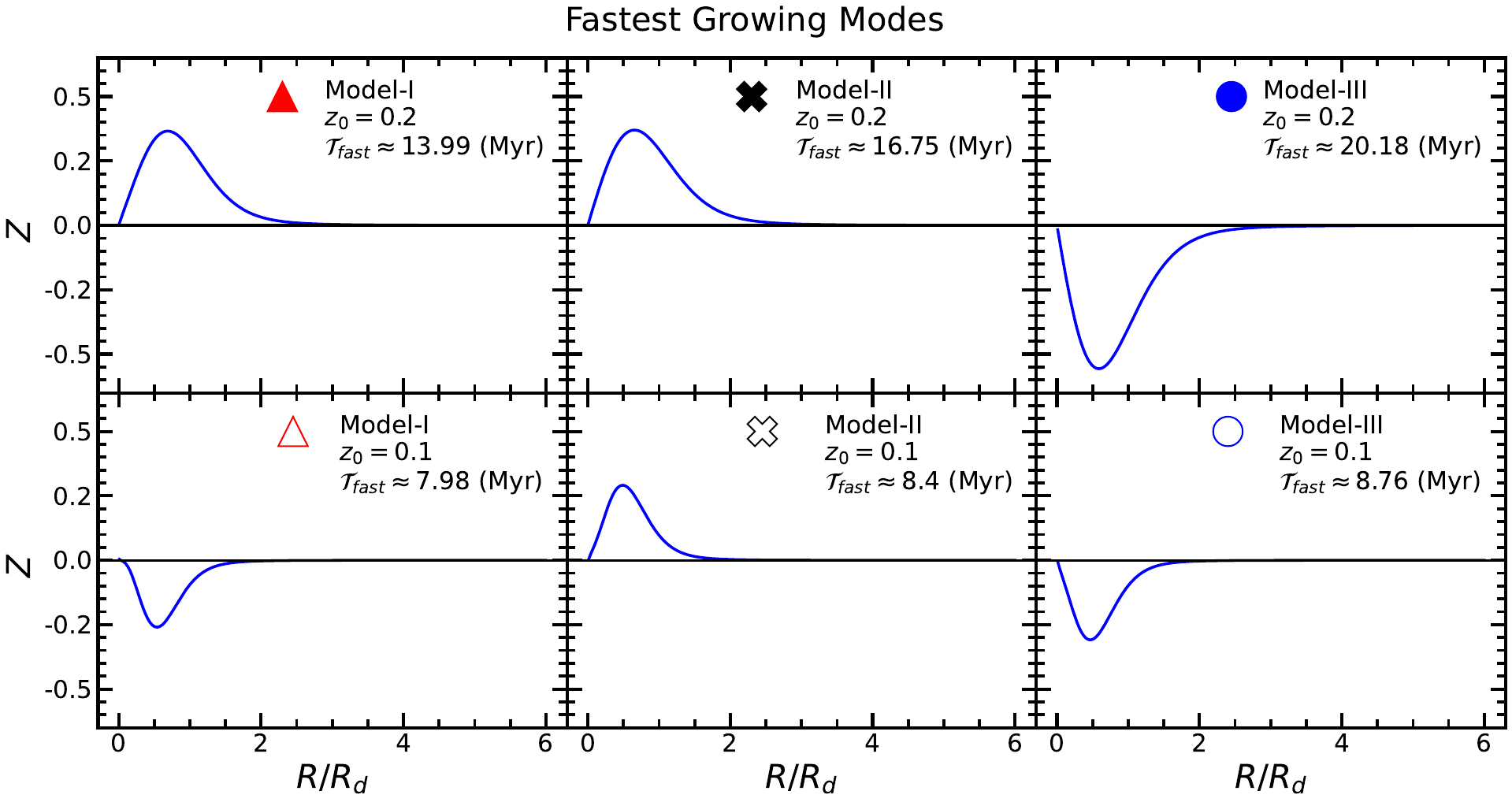}
    \caption{\textbf{Shape of fastest growing mode:} Vertical displacement  $Z(R,t)$ of the fastest growing   unstable modes of the fiducial Model-I, Model-II, and Model-III. The  vertical displacement $Z$ is in the unit of $R_d$. $\mathcal{T}_{fast}$ indicates the timescale of the fastest growing mode, in Myr.}
    \label{fig:Fast_unstable}
\end{figure*}
\section{Dispersion relation and mode wavelength}
\label{sec:WKB}

 In this section, we explore the stability of bending modes using the WKB dispersion relation. The WKB dispersion relation of bending waves depends only on the local properties of the differentially rotating disc. For the $m=1$ mode, the local dispersion relation, assuming thin disc approximation, is given by \citep{2008gady.book.....B}:
 
 \begin{equation}
     [\omega - \Omega (R)]^2 =\nu^2 (R) +2 \pi G \Sigma(R,0) |k|-\sigma_z^2 (R) k^2,
     \label{eqn:wkb}
 \end{equation}
 
\noindent  where $\Sigma(R,0)$ and $k$ are surface density profile and wave number respectively. The self-gravity of the disc stabilizes the bending modes while the vertical velocity dispersion $\sigma_z$ acts as a destabilizing factor. Note that the contribution from the vertical restoring force (first term on the RHS of the above equation) is independent of the scale of the perturbation. In the absence of self-gravity and vertical pressure, an $m=1$ bending wave would propagate like a plane wave with free precession frequencies $\omega = \Omega -\nu$ and $\omega = \Omega +\nu$ (two possible solutions of equation~(\ref{eqn:wkb})).
\par
 Since all the terms on the right-hand side are real, the disc is stable against local perturbation if ${\omega^{\prime}}^2\geq 0$ and unstable otherwise; here $\omega^{\prime}=\omega - \Omega$.
Solving $\frac{d{\omega^{\prime}}^2}{dk} = 0$ provides the critical value of the wave number $k_c$ below that all eigen modes are unstable. Applying this condition in the above equation (\ref{eqn:wkb}) and substituting radial disc surface density $\Sigma(R,0)$ and $\sigma_z^2$,  we obtain the following expression for the critical frequency:

\begin{equation*}
    k_{c} = \frac{\pi G \Sigma}{\sigma_z^2} \text{sign}(k) = \frac{2}{\sqrt{\pi} z_{0}} \text{sign}(k).
\end{equation*}

In other words, the critical wavenumber is entirely determined by the scale height of the disc under this approximation. For $k > k_c$ (short wavelengths $(\lambda<\lambda_c)$) and ${\omega^{\prime}}^2 > 0$,  the solutions are oscillatory; for $k < k_c$ (long wavelengths $(\lambda>\lambda_c)$), the solutions are exponentially growing. The corresponding critical wavelength for unstable modes is $\lambda_c = 2\pi/k_c$.

To calculate the wave number ($k$) and wavelength ($\lambda$) for each unstable mode, we perform a discrete Fourier transform (DFT) on $Z(R)$ in the spatial domain and obtain the one-sided spatial coefficients $H (k)$ \citep{2002nrca.book.....P} as
\begin{equation}
    H (k) = \sum_{r=0}^{N-1} w(r) Z(R) e^{-ikr/N},
\end{equation}

\noindent  where  $r= 0,.....,(N - 1)$.  Here, $w(r)$ is a Gaussian window function with a standard deviation of $N/2^{5/2}$. The window function $w(r)$ is introduced to alleviate spectrum leakage from high frequencies \citep{widrow2014bending}. The discrete wave number is given by 
\begin{equation*}
    k = \frac{t_k}{N\Delta},
\end{equation*}

\noindent where $\Delta$ is sampling interval with $t_k = 0,........,N/2$ such that Nyquist critical limit of wave number corresponding to $t_k = N/2$ is $\frac{1}{2\Delta}$. The power spectrum as a function of wavenumber $k$ is then obtained as

\begin{equation}
    P(k ) =\frac{1}{W} | H(k) |^2, 
\end{equation}

\noindent where $ W = N\sum_{r=0}^{N-1} w(r)$ denotes the window function normalization. The wave number $k$ and corresponding wavelength $\lambda = \frac{2\pi}{k}$ for each unstable mode is obtained at maximum power $P(k)_{max}$.

\begin{figure*}
\centering
\includegraphics[width=0.8\columnwidth]{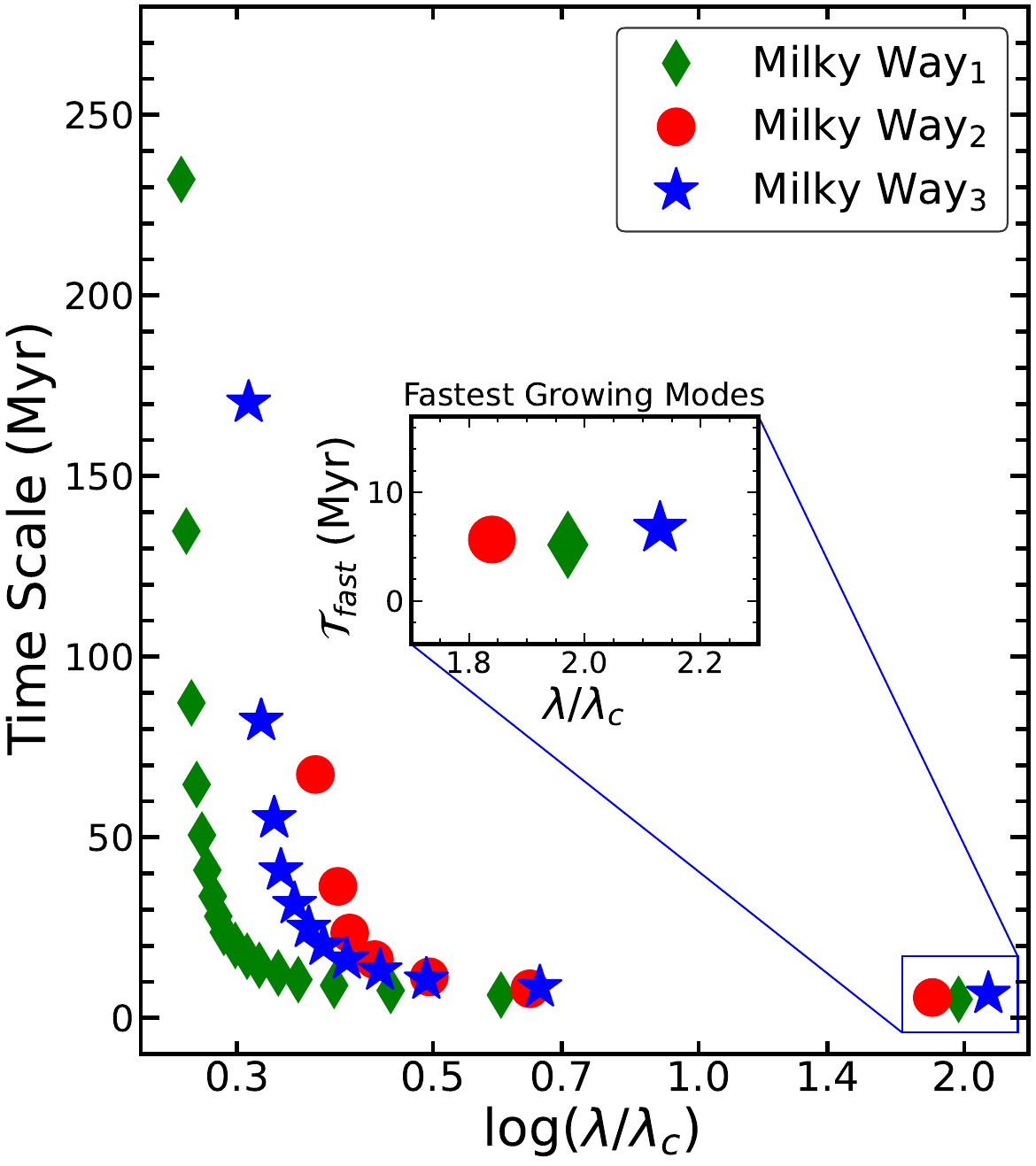}
\hspace{1.5cm}
\includegraphics[width=.55\columnwidth]{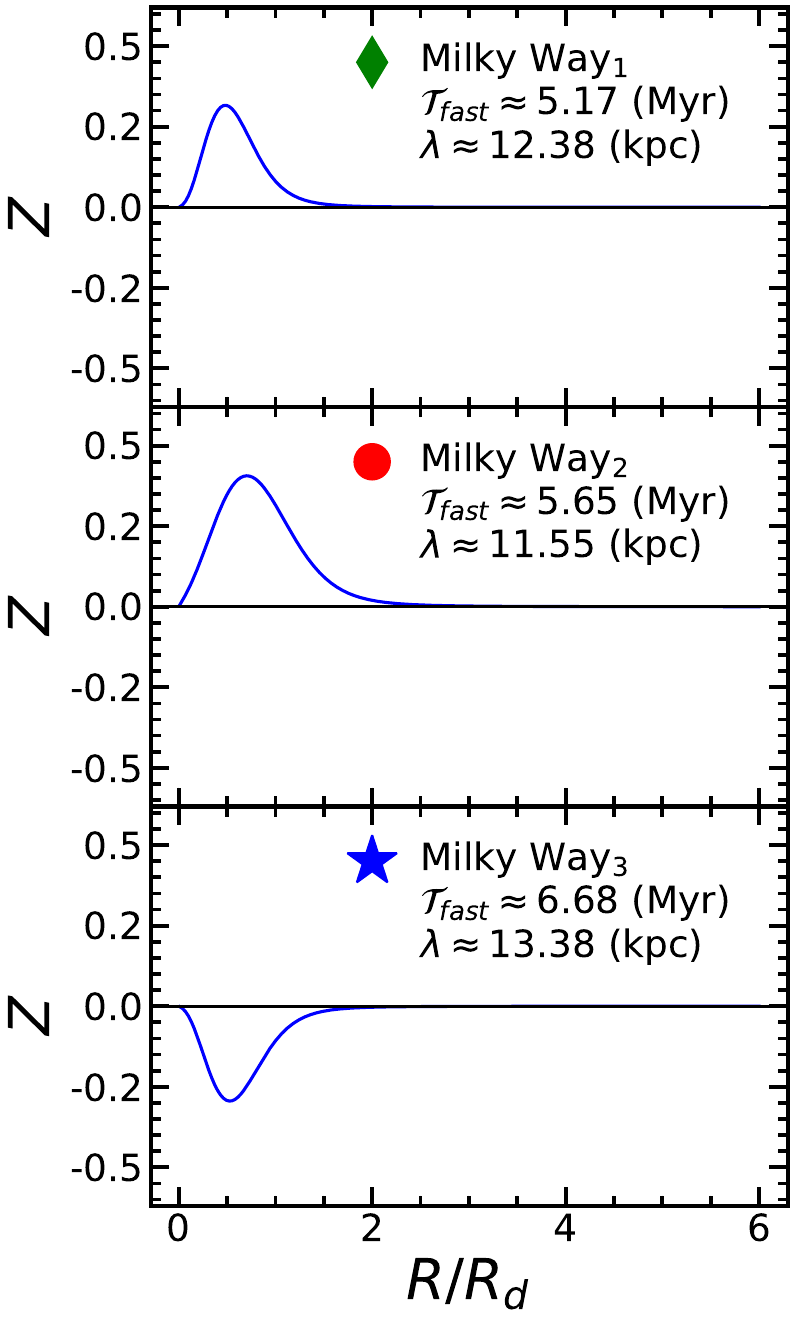}
    \caption{\textbf{Growth time scale and wavelength of modes in Milky Way like galaxies :}  In the left panel,  we  show the growth time scale in Myr of all the unstable modes corresponding to its dimensionless wavelength ($\lambda/\lambda_c$) in the log scale. The inset subplot inside the left panel  shows the fastest growing modes. In the right panel, we show the mode shape of  the fastest growing modes along with wavelengths ($\lambda$) and growth time scales ($\mathcal{T}_{fast}$). Here the unit of wavelength and time scale are kpc and Myr respectively.}
    \label{fig:MW-Timescale_Wavelength}
\end{figure*}

\subsection{Dependence of growth time scale $(\mathcal{T})$ on disc thickness and dark matter halo mass}

 As stated previously, a self-gravitating disc supported by a pressure gradient force possesses unstable eigen modes. The unstable modes have both positive and negative imaginary parts of  eigen frequency ($\omega$). The eigen modes with positive imaginary parts are damped and are of no interest in the present paper. Here, we are mainly interested in the negative imaginary part ($\omega_I$) of the eigenvalues; note both are the same in magnitude since eigen values appear in complex conjugate pair. The modulus of $\omega_{I}$ refers to the growth rate while its reciprocal measures the growth time scale $(\mathcal{T} = 1/|\omega_{I}|)$ of the unstable modes. The mode with the shortest growth time scale is the one that grows fastest among all the unstable modes. In this paper, the shortest growth time scale is denoted by $\mathcal{T}_{fast}$.

\begin{table}
\caption{\textbf{Wavelength and growth time scale:} This table give the real ($\omega_{_R}$) and imaginary ($\omega_{_I}$) parts of eigen frequencies and corresponding wavelength ($\lambda$) and  growth time-scale $\mathcal{T}_{fast}$ of the fastest growing unstable modes obtained from Model-I, II, and III,  in presence of vertical pressure.   The wavelength and  time scale $\mathcal{T}_{fast}$ given here are in the unit of $R_d$ and $\sqrt{R_d^3/GM_d}$ respectively.} 
\label{table: wavelength & time scale}
\centering
\begin{tabular}{c c c c c c c}     
\toprule\toprule
$z_{_0}$ &  $M_h/M_d$ &$\omega_{_R}$  & $|\omega_{_{I}}|$  & $\mathcal{T}_{fast}$ &$\lambda_c $& $\lambda$ \\
\hline
   0.1 &   	  3	    &	0.77 & 1.95 & 0.51 & 1.11 & 4.70 \\

       &    	10	   &	0.91 &  1.85 & 0.53 &  & 5.37 \\

       &   	15	   &	1.01 &   1.77 & 0.56 &   & 6.27\\
\midrule
 0.2  &        3	    &	 0.51  &     1.11 & 0.89 &  1.67&  12.54 \\

      &    	10    &	   0.71   &   0.93 & 1.07 &  & 12.54\\

      &    	15    &	  0.82    &  0.77  & 1.29 &  & 18.81 \\
\bottomrule
\end{tabular}
\end{table}

 In Table.\ref{table: wavelength & time scale}, we have given the  growth time-scale $\mathcal{T}_{fast}$ and wavelength of the fastest unstable modes of Model-I, Model-II, and Model-III of the disc having different scale heights $z_0 =0.1$, and $z_0=0.2$. {\it All the fastest growing modes have the longest wavelengths among all the unstable modes}. These wavelengths are found to be significantly longer than the critical wavelength ($\lambda_c$) for both the disc scale height values as expected from the WKB dispersion relation. These unstable modes are the ones with the smallest number of radial nodes. It is also interesting to note that the wavelength of the fastest mode increases with the increasing disc thickness. For a thin disc ($z_0 =0.1$), we obtain the wavelength comparable to the size of the disc ($\sim 6R_d$). However, for the larger thickness disc (e.g. see Table. \ref{table: wavelength & time scale} for $z_0=0.2$) the wavelength is significantly longer than the size of the disc.   
 
We next calculate the growth time scale of the fastest mode. In all three models, we see that the values of $\mathcal{T}_{fast}$ increase when the values of the disc scale heights $z_0$ increase (see Table.\ref{table: wavelength & time scale}). For Model-I, we get $\mathcal{T}_{fast}= 0.51$, and $\mathcal{T}_{fast}= 0.89 $ for $z_0=0.1$ and $z_0=0.2$ respectively. For Model-II, the values of $\mathcal{T}_{fast}$ having $z_0=0.2$ and $0.1$ are 1.07 and 0.53 respectively. For $z_0 = 0.2$, the Model-III  has only one unstable mode with a growth time scale of $\mathcal{T}_{fast} =1.29$, while for $z_0 = 0.1$, we get  the value of $\mathcal{T}_{fast} = 0.56$. The general trend is that the unstable mode in a thicker disc grows slowly whereas it grows faster in a thin disc. In thinner discs, the existence of non-zero vertical velocity dispersion allows the disc to buckle and grow more quickly surpassing the combined effect of restoring forces due to the self-gravitating disc and dark matter halo. \textit{In other words, we can conclude that thin discs are more unstable than thick discs in the light of vertical pressure gradient force.}
 
Further, we explore the effect of dark matter halo mass on the growth time scale of unstable modes. From the numerical values of $\mathcal{T}_{fast}$ (see Table.\ref{table: wavelength & time scale}) for the three models, we observe the following general trends of unstable modes in the disc. (1) Growth time scale of the unstable mode is larger in the presence of a more massive dark matter halo resulting in a more stable disc. (2) In the presence of low mass dark matter halo, unstable modes in the disc grow faster.

 In the left panel of Fig.\ref{fig:timescale}, we show a plot of time scales ($\mathcal{T}$) in Gyr  with respect to dimensionless wavenumbers $k/k_{c}$ of  unstable modes of three models using fiducial disc mass  $M_d = 3 \times10^{10}$ M$_\odot$ and scale radius $R_d=3.2$ kpc.   Different shapes represent the  unstable modes with different wavenumbers and growth time scales  for the models. All the unstable modes are found below the critical wave number $k_c$ i.e. $k/k_c<1$. In the right panel of Fig.\ref{fig:timescale}, we show that  the unstable mode having  the smallest wavenumber (longest wavelength) has the shortest growth time scale in Myr. In Fig.\ref{fig:Fast_unstable}, we show the vertical displacement of fastest growing unstable eigen modes along with the growth time scales.

\subsection{Bending instability in Milky Way like galaxies} 
\label{sec6}

For a better understanding of bending instability in realistic galaxies, we apply our theoretical model to Milky Way like galaxies. Each of these models are characterized by a set of parameters taken from recent literature, see Table.\ref{table:MW} for details. We adopted these parameters for the Galaxy and dark matter halo because they are the best-fitting parameters given by \cite{1998A&A...330..953M,2010AN....331..731K,2011MNRAS.414.2446M,2013ApJ...779..115B}. We estimate the growth time scale and corresponding wavelength of the fastest unstable mode for each of these models. 

In the left panel of Fig.\ref{fig:MW-Timescale_Wavelength}, we show the dimensionless wavelength $\lambda/\lambda_c$ in log scale and growth time scale of the unstable modes that arise in the three models of the Milky Way. We obtain a few unstable modes for all the Galaxy models. The critical values of wavelengths of all three models Milky Way$_1$, Milky Way$_2$, and Milky Way$_3$ are 1.67~kpc, 2.22~kpc, and 1.78~kpc respectively. All the unstable modes have wavelengths above the critical wavelength i.e. $\lambda/\lambda_c>1$. We estimate the wavelength of unstable modes, particularly the fastest-growing modes. The wavelengths of fastest growing modes are found to be $\lambda = 12.4$~kpc,  11.5~kpc,  and 13.4~kpc for the Milky Way$_1$, Milky Way$_2$, and Milky Way$_3$ respectively. The mode shapes of these unstable modes are shown in the right panel of Fig.\ref{fig:MW-Timescale_Wavelength}. It is interesting to note that wavelengths of the fastest-growing modes are somewhat found to be as comparable to the radius of the discs. Further, the estimated growth time scales $\mathcal{T}_{fast}$ of the fastest modes are found to be 5.2~Myr, 5.5~Myr, and 6.7 Myr respectively.

\section{Discussion and Conclusions} 
\label{sec7}

The bending instability of a rotating self-gravitating galactic disc has attracted some of the best minds in galaxy dynamics and continues to be a subject of profound interest amongst dynamicists. A recent surge in the activities is due to the availability of high-quality data on stellar motion (6D phase space) in the Milky Way by the Gaia satellite \citep{2018A&A...616A..11G}. Gaia has made it possible to identify and investigate bending modes with minute details in the Galaxy \citep{widrow2012galactoseismology, widrow2014bending, xu2015rings, 2018MNRAS.478.3809S, 2018Natur.561..360A, 2018MNRAS.481L..21P, bennett2019vertical}. Not only our Galaxy but bending waves are also characterized recently in a number of external galaxies via line-of-sight kinematics of ionized gas (H$\alpha$),  neutral hydrogen gas,  \citep{Sanchez+2015, 2022MNRAS.513.3065N} as well as corrugated dust pattern \citep{2020MNRAS.495.3705N}. These observations have been a clear motivation to revisit this age-old problem. However, we do not attempt to explain any of these observations depicting bending modes in this work but rather attempt to provide a more detailed picture of the bending instability from an analytic point of view.

In this work, we reaffirm that in a razor-thin disc, in the absence of vertical pressure gradient force, all eigenmodes in the spectrum have real eigenvalues \citep{HunterToomre1969, Sellwood1996, saha2006generation, Chequers+2018}. The stable eigen modes are, generally, parts of two main continuum branches: slow modes continuum with negative eigen frequencies ($\omega< 0$) and fast modes continuum with positive eigen frequencies ($\omega> 0$). In addition to the two continuum modes, a razor-thin disc might support a discrete, stable large-scale eigen mode. Such a discrete mode has been a cornerstone of numerical work in several previous investigations  \citep{HunterToomre1969, Mathur1990, binney1992warps, Sellwood1996}. A discrete bending mode would behave like an oscillating pattern in the disc without any damping and growing. If found, such a discrete mode could be a viable solution, as envisaged previously, for the observed warp in many disc galaxies
\citep{1998A&A...337....9R, 2003A&A...399..457S, AnnPark2006}. However, as shown by the previous numerical works of Hunter, Toomre, and Sellwood, a discrete mode existed only when the disc was sharply truncated. Although theoretical work brought the discrete mode picture alive \citep{Mathur1990, Louis1992}, they were for a simplified system, and naturally a need for an extensive numerical analysis on realistic models of galaxies was in place. For a smoothly truncated exponential disc, the existence of discrete $m=1$ bending mode was shown via numerical work in \cite{saha2006generation}. In the current work, we perform an extensive search for the discrete bending  in more realistic models of disk galaxies by including dark matter and vertical pressure in the dynamical equation.

Our current analysis is based on Binney's logarithmic dark matter halo potential, producing a flat rotation curve. It is interesting to see that the number of unstable modes in the P-gap decreases as the halo mass increases with respect to the disc. In the absence of disc self-gravity, it is the halo that supplies the restoring force against the destabilizing pressure force. In that spirit, it might be useful to know how the bending instability arises in the presence of a dark matter halo with different density profiles such as the Navarro Frank White (NFW) profile \citep{1996ApJ...462..563N} which is common
in cosmological simulations of structure formation \citep{boylan2009resolving}. Since for the NFW halo, the density falls as $r^{-3}$, the local restoring force due to the halo would be lowered as compared to the logarithmic halo (where $\rho \propto r^{-2}$) in the outer parts and if the disc self-gravity is not significant, an NFW halo might promote stronger bending instability compared to the logarithmic halo. We plan to explore this in detail in a future paper. 

Our models of disc galaxies have one disc component. Real galaxies, e.g., our Milky Way and several external galaxies have both a thin and thick disc \citep{2014ApJ...787...24B, 2017MNRAS.465.3784B, 2023arXiv230304171H}. It would be insightful, how the thick disc component would affect the bending instability. Finally, our current analysis is restricted to only the $m=1$ bending mode. Higher order bending modes or corrugations (seen in stars, dust, or gas) are also known in several galaxies \citep{Quiroga1974, 1982ApJ...259L..63K, saha2006generation, Levine+2006, 2020MNRAS.495.3705N}. Future exploration of higher order bending modes in gravitationally coupled discs of stars and gas would help in getting a complete picture of bending instabilities in a realistic disc galaxy. We draw the following main conclusions based on this work:
\begin{itemize}

     \item In a smoothly truncated, exponential  razor-thin disc, we reaffirm that all the $m=1$ bending modes are stable. Such disc might support a stable discrete mode in the P-gap, describing  classic integral-sign warps  with a purely oscillatory nature. The general properties of such  discrete mode are sensitive to the halo core radius. In particular, the modes cannot maintain the S-shaped nature inside a halo with large core radii and ceases to exist in the P-gap at very small core radii.
    
    \item In a realistic galaxy model, the vertical pressure of the disc excites unstable modes in the P-gap as well as outside the P-gap, as expected from the WKB relation. Such a disc also supports discrete stable long-lived modes in the P-gap. We show that the increasing halo mass first stabilizes the unstable modes in the P-gap and then the modes outside the P-gap, resulting in overall a smaller number of unstable modes at a very high halo mass. On the other hand, the increasing halo core radius causes more unstable modes to arise in the eigen spectrum. In other words, the overall instabilities of the disc are highly governed by the dark matter halo.

    \item Our numerical analyses show that in a thin disc, the vertical pressure excites a greater number of unstable modes outside the P-gap as well as in the gap than its thicker counterpart.

    \item  Using WKB dispersion relation and discrete Fourier  transform (DFT), we show that all the unstable modes have wavelengths above the critical value of wavelength. Our analyses reveal that unstable modes with the longest wavelength have the fastest growth time scale. The growth time scale is found to be affected by the halo mass and disc thickness. In a low halo mass, the unstable modes grow faster than in a massive halo. For a fixed halo mass, the models grow faster in a thinner disc.
    
    \item For Milky Way-like galaxies, the wavelength of the fastest growing mode is found to lie approximately within the range of $\lambda \simeq  11 - 13$~kpc; comparable to the radius of the stellar disc, and the growth time scale lies within the range of  $5 - 7$~Myr.

\end{itemize} 

\section*{Acknowledgements}

The authors acknowledge Rajiv Gandhi University, Arunachal Pradesh, and Inter-University Centre for Astronomy and Astrophysics (IUCAA), Pune for providing local hospitality and computational facilities to carry out this research work. The Python packages Numpy, Scipy, and Astropy   are used in  numerical calculations and analysis. Further, Sagar S. Goyary acknowledges the UGC-CSIR (Govt. of India) for the Senior Research Fellowship to support financially during the period of the present work.

\section*{Data Availability}
The data underlying this article will be shared on reasonable request to the corresponding author.

\bibliographystyle{mnras}
\bibliography{Reference} 
\label{lastpage}
\end{document}